\documentclass[journal]{IEEEtran}
\usepackage{cite}
\usepackage{amsmath}
\usepackage{amsthm}
\usepackage{graphicx}
\usepackage{algorithm}
\usepackage{setspace}
\usepackage{caption}
\usepackage{subcaption}
\usepackage{varwidth}
\usepackage[noend]{algpseudocode}
\usepackage{setspace}
\usepackage{hyperref}
\usepackage{amssymb}
\usepackage[noend]{algpseudocode}
\usepackage{array}
\usepackage{blindtext}
\newtheorem{remark}{Remark}
\newtheorem{assumption}{Assumption}
\DeclareMathOperator*{\argmin}{arg\,min}
\usepackage{xcolor}
\begin{document}
\title{Data-Driven Strategies for Hierarchical Predictive Control in Unknown Environments}
\author{Charlott~Vallon
        and~Francesco~Borrelli,~\IEEEmembership{Fellow,~IEEE}% <-this % stops a space
\thanks{
This research was partially funded by the National Physical Science Consortium, the National Institute of Standards and Technology,
the Office of Naval Research grant ONR-N00014-18-1-2833 and by the National Science Foundation grant NSF-1931853.}
\thanks{C. Vallon and F. Borrelli are with the Department of Mechanical Engineering, University of California, Berkeley, Berkeley, CA 94720.}% <-this % stops a space}
}

% \markboth{Transactions on Automation Science and Engineering}%
% {Shell \MakeLowercase{\textit{et al.}}: Bare Demo of IEEEtran.cls for IEEE Journals}

\maketitle
% Safe Data-Driven (Predictive) Control in Changing Environments via Strategies and Hierarchies

\begin{abstract}
% We describe a hierarchical learning architecture for safe predictive control in unknown environments. 
% We consider a constrained nonlinear dynamical system and assume the availability of state-input trajectories solving control tasks in different environments. 
% A parameterized environment model generates state constraints specific to each task, which are satisfied by the stored trajectories.
% Our goal is to find a safe and high-performance data-driven policy for a new task in an unknown environment. 
% From stored data, we learn strategies in the form of target sets in a reduced-order state space.
% These strategies are applied to the new task in real-time using a local forecast of the new environment, and the resulting output is used as a terminal region by a low-level model predictive controller.
% We show how to \emph{i)} design the target sets from past data and then \emph{ii)} incorporate them into a model predictive control scheme with shifting horizon that ensures safety of the closed-loop system when performing the new task. 
% We prove the feasibility of the resulting control policy, and verify the proposed method in robotic path planning, autonomous racing, and computer game applications.

This article proposes a hierarchical learning architecture for safe data-driven control in unknown environments. 
We consider a constrained nonlinear dynamical system and assume the availability of state-input trajectories solving control tasks in different environments. 
In addition to task-invariant system state and input constraints, a parameterized environment model generates task-specific state constraints, which are satisfied by the stored trajectories.
Our goal is to use these trajectories to find a safe and high-performing policy for a new task in a new, unknown environment. 
We propose using the stored data to learn generalizable control strategies.
At each time step, based on a local forecast of the new task environment, the learned  strategy consists of a target region in the state space and input constraints to guide the system evolution to the target region.
These target regions are used as terminal sets by a low-level model predictive controller.
We show how to \emph{i)} design the target sets from past data and then \emph{ii)} incorporate them into a model predictive control scheme with shifting horizon that ensures safety of the closed-loop system when performing the new task. 
We prove the feasibility of the resulting control policy, and apply the proposed method to robotic path planning, racing, and computer game applications.

\end{abstract}


\section{Introduction}\label{sec:introduction}

\IEEEPARstart{C}{lassical} Iterative Learning Controllers (ILCs) aim to improve a system's closed-loop reference tracking performance at each iteration of a repeated task while rejecting periodic disturbances \cite{AA, shen2019survey}. In classical ILC, the controller uses tracking error data from previous task iterations to better track the provided trajectory during the current iteration.
These methods have been used in a variety of applications, ranging from quadrotor control \cite{ilcUAV} to additive manufacturing \cite{armstrong2021multi}, robotics \cite{hofer2019iterative}, and ventricular assist devices \cite{ketelhut2019iterative}.
Recent work has also explored reference-free ILC for applications whose goals are better defined through a performance metric, such as autonomous racing or harvesting wind energy \cite{rosolia2017autonomousrace, melanie, vermillion}. 
%The controller again uses previous iteration data to improve closed-loop performance with respect to the chosen performance metric.

Because ILC policies are trained to attenuate constant or repeated task-specific disturbances, the learned ILC policy is generally not cost-effective, or even feasible, if the task environment or reference trajectory changes \cite{rosolia2017autonomousrace, vallon2019task}. Many ILC  approaches have been proposed for improving ILC performance in these situations, including representing tasks in terms of basis functions \cite{van2016optimality, alleyne2010basis}, enforcing sparsity in the ILC policy \cite{rojas2017}, or training neural networks to predict the task-specific ILC input given an environment observation \cite{tomi2020neural, patan2020neural}.

In the Artificial Intelligence and Reinforcement Learning communities, the ability to generate a control policy which performs well under different environment conditions is a common challenge, often referred to as generalization or transfer learning \cite{transferlearning, weiss2016survey, zhuang2019comprehensive}.
Model-free approaches typically focus on minimizing a policy's performance loss between environments \cite{finnMetaRL2018, levine2018hierarchical}, rather than guaranteeing feasibility of the policy in a new environment.

Here, we focus specifically on methods that use stored trajectory data from previous tasks in order to find feasible and effective policies for a new task.
Methods that do guarantee feasibility (at least with high likelihood) are generally model-based \cite{bertsimas2018voice, hewing2019cautious, stolle, berenson, pereida2018data}, and approaches have been proposed for autonomous vehicles \cite{zhi2019octnet} and robotic manipulation applications \cite{probabilisticprimitives, fitzgerald, vallon2019task}.
In \cite{pereida2018data}, a static map is learned between a given reference trajectory and the input sequence required to make a linear time-invariant system track that reference trajectory. Once learned, this map can be used to determine an input sequence that lets the linear system track a new reference trajectory. 
Many other strategies propose maintaining trajectory libraries \cite{berenson, 6, liu2009standing, 7}, and adapting the stored trajectories online to the new constraints of the changed tasks, which can be both time-consuming and computationally expensive.
%The authors in \cite{BG} propose running a desired planning method in parallel with a retrieve and repair algorithm that adapts reference trajectories from previous tasks to the constraints of a new task. Retrieve and repair was shown to decrease overall planning time, but requires checking for constraint violations at each point along a retrieved trajectory.
%In \cite{6}, environment features are used to divide a task and create a library of local trajectories in relative state space frames. These trajectories are then pieced back together in real-time according to the features of the new task environment.
%A trajectory library built using differential dynamic programming is used in \cite{liu2009standing} to design a controller for balance control in a humanoid robot. At each time step, a trajectory is selected from the library based on current task parameter estimates and a k-nearest neighbor selection scheme. A similar method is explored in \cite{7}, where differential dynamic programming is combined with receding horizon control. While these methods can decrease planning time, they verify or interpolate saved trajectories at every time step, which can be inefficient and unnecessary.

This article proposes a novel data-driven method for tackling a simple abstraction of the changing environment control problem.
We consider availability of state-input trajectories which solve a set of $n$ control tasks $\{\mathcal{T}^1$,\ldots,$\mathcal{T}^n\}$.
%A successful execution of the $i^\mathrm{th}$ control task $\mathcal{T}^i$ is defined as a trajectory of states and inputs evolving according to a nonlinear difference equation, satisfying both system state and input constraints as well as constraints imposed by the environment.
In  each of the $n$ control tasks the system model, system constraints, and objective are identical. However, a parameterized  environment descriptor function ${\bf{\Theta}}^i$ generates a set of environment state constraints specific to each task $\mathcal{T}^i$. 
Our goal is to use stored executions of previous control tasks in order to find a successful execution of a new task
$\mathcal{T}^{n+1}$ in a new environment described by ${\bf{\Theta}}^{n+1}$.

This paper presents a hierarchical predictive learning architecture to solve such a problem. We use stored task data to design sets in a reduced-dimension state space that can be interpreted as high-level strategies learned from previous control tasks, and are used as waypoints to track by a low-level model predictive control (MPC).
%The word \textit{strategy} in this paper will refer to such target sets.
In this paper, we:
\begin{enumerate}
    \item propose interpreting strategies as sets in reduced-dimension state space, referred to as ``strategy sets," 
    \item show how to design strategy sets from past data,
    \item demonstrate how to incorporate the strategy sets into an MPC, 
    \item prove that the proposed method will lead to a successful execution of the new control task, and
    \item verify our method in three different applications.
    %\item provide a discussion on the benefits using a hierarchical framework can offer for control generalizability, modularity, understanding, and safety. 
\end{enumerate}
%Our approach differs from other proposed hierarchical approaches \cite{ugo2020unified, levine2018hierarchical} in that we... blah, we are safe, general, but use models where appropriate so as to guarantee safety, and argue that human input is good? 

This paper builds on the author's work published in~\cite{vallon2020data} and includes a novel  method for relaxing the feasibility guarantees, shows how the algorithm can be combined with machine learning techniques, and provides a more thorough experimental evaluation.

The remainder of this paper is organized as follows. Section~\ref{sec:problem_formulation} formalizes the background information and problem statement. Section~\ref{sec:hpl} introduces the Hierarchical Predictive Learning Control Framework, which is detailed further in Secs.~\ref{sec:strategyfinder}-\ref{sec:algorithm}. We conclude with a proof of the feasibility of our proposed policy and three example applications. 
We conclude our paper with a discussion in Sec.~\ref{sec:discussion}.

\section{Problem Formulation}\label{sec:problem_formulation}

We consider a discrete-time system with dynamical model
\begin{align}\label{eq:VehicleModel}
    x_{k+1} &= f(x_k, u_k),
\end{align}
subject to system state and input constraints
\begin{align}\label{eq:VehicleModelConstr}
    x_k \in \mathcal{X},~ u_k \in \mathcal{U}.
\end{align}
The vectors $x_k \in \mathbb{R}^{n_x}$ and $u_k \in \mathbb{R}^{n_u}$ collect the states and inputs at time $k$.

\subsection{Tasks}
The system (\ref{eq:VehicleModel}) solves a series of $n$ control tasks $\{\mathcal{T}^1, \dots, \mathcal{T}^n\}$. Each control task $\mathcal{T}^i$ is defined by the tuple
\begin{equation}
    \mathcal{T}^i = \{\mathcal{X}, \mathcal{U}, \mathcal{P}^i, {\bf{\Theta}}^i \},
\end{equation}
where $\mathcal{X}$ and $\mathcal{U}$ are the system state and input constraints (\ref{eq:VehicleModelConstr}). $\mathcal{P}^i \subset \mathcal{X}$ denotes the task target set the system (\ref{eq:VehicleModel}) needs to reach in order to complete task $\mathcal{T}^i$, and ${\bf{\Theta}}^i$ is an environment descriptor function, formalized below. 
%\textcolor{red}{Formalize what a task is. Does it have an end? Infinite vs. finite time; is it defined by a task length, or by a goal set, or what?}

\subsection{Task Environments}
\label{ssec:tasksandexecutions}
%\textcolor{red}{This has to be updated to match the new formulation from the quals presentation.}
The $n$ solved control tasks $\{\mathcal{T}^1, \dots, \mathcal{T}^n\}$ take place in different task environments, parameterized by different environment descriptor functions $\{{\bf{\Theta}}^1, \dots, {\bf{\Theta}}^n\}$. In each control task the system model (\ref{eq:VehicleModel}) and constraints (\ref{eq:VehicleModelConstr}) are identical. However, the environment descriptor function generates additional task-specific environmental state constraints. 

For each task $\mathcal{T}^i$, the environment descriptor function ${\bf{\Theta}}^i$ maps the state $x_k$ at time $k$ to a description of the local task environment
${\bf{\Theta}}^i(x_k)$. The set of states satisfying the environmental constraints imposed by this local task environment are denoted by $E({\bf{\Theta}}^i(x_k))$. Note that in this paper we only consider state-dependent, but time-invariant, functions ${\bf{\Theta}}^i$.

We write the joint system and environment constraints as 
\begin{align}\label{eq:taskconstraints}
    x_k \in \mathcal{X}({\bf{\Theta}}^i(x_k)) =  E({\bf{\Theta}}^i(x_k)) \cap \mathcal{X}.
\end{align}
% where
% \begin{align}
%     \mathcal{X}({\bf{\Theta}}^i(x_k)) =  E({\bf{\Theta}}^i(x_k)) \cap \mathcal{X}. \nonumber
% \end{align}
%Since ${\bf{\Theta}}^i(x_k)$ may be time-varying, the combined constraints (\ref{eq:taskconstraints}) may also be state- and time-dependent.
For notational simplicity, wherever it is obvious we will drop the state dependence and refer to the combined system and environment constraints (\ref{eq:taskconstraints}) as $\mathcal{X}({\bf{\Theta}}^i)$. 

\begin{figure}
     \centering
     \begin{subfigure}[b]{\columnwidth}
         \centering
         \includegraphics[width=0.7\columnwidth]{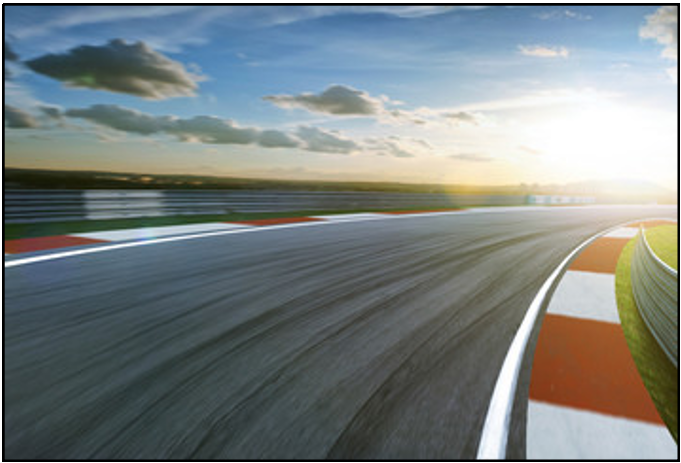}
         \caption{The environment descriptor function ${\bf{\Theta}}$ gives a local description of the task environment, such as a camera image of the track at a particular location.}
         \label{fig:roadpic}
     \end{subfigure}
    \par\bigskip
     \begin{subfigure}{\columnwidth}
         \centering
         \includegraphics[width=0.7\columnwidth]{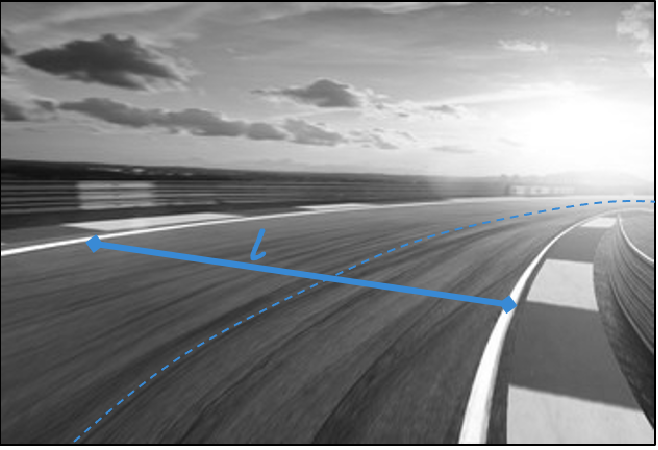}
         \caption{From this description, local environment-dependent constraints such as lane boundaries the vehicle must remain within to successfully complete the task, can be determined.}
         \label{fig:road_cv}
     \end{subfigure}
      \par\bigskip
    \caption{ }
    \label{fig:road}
\end{figure}

Examples of local task environment descriptions ${{\bf{\Theta}}^i(x_k)}$ include camera images, the coefficients of a polynomial describing a race track's lane boundaries, or simple waypoints for tracking. 
Consider, for example,  an autonomous racing task where a car has to drive around a track as quickly as possible, so that the task target set $\mathcal{P}$ contains the finish line. 
Here, the vehicle-specific state and input constraints (\ref{eq:VehicleModelConstr}) may include acceleration and steering limits imposed by the construction of the vehicle itself. The environment descriptor function ${\bf{\Theta}}^i$ may map a state along the track to a frontview camera image of the track at that point in space and time (Fig.~\ref{fig:roadpic}). From this camera environment description, we can then extract the additional environmental constraints $E({\bf{\Theta}}^i(x_k))$, such as lane boundaries the vehicle needs to remain in, which are dictated by the environment rather than the vehicle (Fig.~\ref{fig:road_cv}).

% Note that ${\bf{\Theta}}(x_k, k)$ would more likely map to a lower-dimensional encoding of a camera image or similar environment description such as racetrack curvature.

\subsection{Environment Forecasts}
The control architecture proposed in this paper relies on forecasts of the task environment, which can be obtained using the environment descriptor function.
At time $k$, we can use (\ref{eq:VehicleModel}) to predict the system state across a horizon $N$, 
\begin{align*}
    \hat{x}^i_{k:k+N} = [\hat{x}^i_{k}, \hat{x}^i_{k+1}, \dots, \hat{x}^i_{k+N}],
\end{align*}
and evaluate the environment descriptor function along this forecast state trajectory. This provides a forecast of both the upcoming task environment ${\bf{\Theta}}^i(x_k)$ and  environmental constraints $E({\bf{\Theta}}^i(x_k))$:
\begin{align}\label{eq:env_forecast}
    \theta^i_{k:k+N} = [{\bf{\Theta}}^i(\hat{x}^i_{k}),\dots, {\bf{\Theta}}^i(\hat{x}^i_{k+N})].
\end{align}

In the context of the autonomous racing example, we can evaluate ${\bf{\Theta}}^i$ at forecasts of the system state in order to get predictions for the upcoming racetrack curvature, which can be evaluated from the camera images (Fig.~\ref{fig:road}). 

\subsection{Task Executions}
A feasible execution of the task $\mathcal{T}^i$ in an environment parameterized by ${\bf{\Theta}}^i$ is defined as a pair of state and input trajectories
    \begin{align}\label{eq:iterationvecs}
\mathrm{Ex}(\mathcal{T}^i, {\bf{\Theta}}^i) & = [{\bf{x}}^i, {\bf{u}}^i] \\ 
    {\bf{x}}^i & = [x^i_{0},x^i_{1},...,x^i_{D^i}], ~x^i_k \in \mathcal{X}({\bf{\Theta}}^i)~~ \forall k \in [0, {D^i}], \nonumber\\ & ~~~~~~~~~~~~~~~~~~~~~~~ x^i_{D^i} \in \mathcal{P}^i,\nonumber\\
    {\bf{u}}^i & = [u^i_0,u^i_1,...,u^i_{D^i}], ~u^i_k \in \mathcal{U}~~~~~~~~ \forall k \in [0, D^i],\nonumber
    \end{align}
where ${\bf{u}}^i$ collects the inputs applied to the system (\ref{eq:VehicleModel}) and ${\bf{x}}^i$ is the resulting state evolution. $D^i$ is the duration of the execution of task $\mathcal{T}^i$. The final state of a feasible task execution, $x^i_{D^i}$, is in the task's target set $\mathcal{P}^i \subset \mathcal{X}({\bf{\Theta}}^i)$.

In the racing task, feasible executions would be vehicle state and input trajectories obeying the lane boundaries described by the environment descriptor function.

\subsection{Problem Definition}

Given a dynamical model (\ref{eq:VehicleModel}) with state and input constraints (\ref{eq:VehicleModelConstr}),~(\ref{eq:taskconstraints}), and a collection of feasible executions (\ref{eq:iterationvecs}) that solve a series of $n$ control tasks, $\{\mathrm{Ex}(\mathcal{T}^1, {\bf{\Theta}}^1), ..., \mathrm{Ex}(\mathcal{T}^n, {\bf{\Theta}}^n) \}$, our aim is to find a data-driven control policy $\pi(x)$ that results in a feasible and high-performance execution of a new task in a new environment: $\mathrm{Ex}(\mathcal{T}^{n+1}, {\bf{\Theta}}^{n+1})$.

\begin{remark}
\textit{Remark:} For notational simplicity, we write that the collected data set contains one execution from each task.
However, in practice one may collect multiple executions of each task $\mathcal{T}^i$. In this case, the executions can simply be stacked, and the procedure proposed in this paper can proceed as described.
\end{remark}

If feasibility were the only goal, one approach would be to use a robust, conservative controller. 
However, we would also like to solve the new task well.
Thus, in addition to satisfying the new environment constraints, the execution should try to minimize a desired objective function $J({\bf{x}}^{n+1}, {\bf{u}}^{n+1})$. 
We do not explicitly take into account a cost function $J$ in this paper. Instead, we assume that the stored executions from previous tasks $\mathcal{T}^i$ were collected using control policies that aimed to minimize the same cost function $J({\bf{x}}^{i}, {\bf{u}}^i)$.
Thus the desired objective function $J$ is implicitly associated with the stored data.
If this assumption does not hold, the method proposed in this paper will still provide a feasible trajectory.
An approach which explicitly computes and compares the costs of different trajectories for when this assumption does not hold is presented in \cite{vallon2019task}.

\section{Hierarchical Predictive Learning Control}\label{sec:hpl}
We propose a data-driven controller that uses stored executions from previous tasks to find a feasible policy for a new task in a new environment.
Instead of simply adapting the stored executions from previous tasks to the changed environmental constraints of the new task, we learn generalizable and interpretable \textit{strategies} from past task data, and apply them to the new task. Our approach is inspired by how navigation tasks are typically explained to humans, who can easily generalize their learning to new environments by learning strategies. 

\subsection{A Motivating Example}\label{ssec:racerules}

Consider learning how to race a vehicle around a track. If a human has learned to race a vehicle by driving around a single track, they can easily adapt their learned strategy when racing a new track.

\begin{figure}[h!]
    \centering
    \includegraphics[width=0.9\columnwidth]{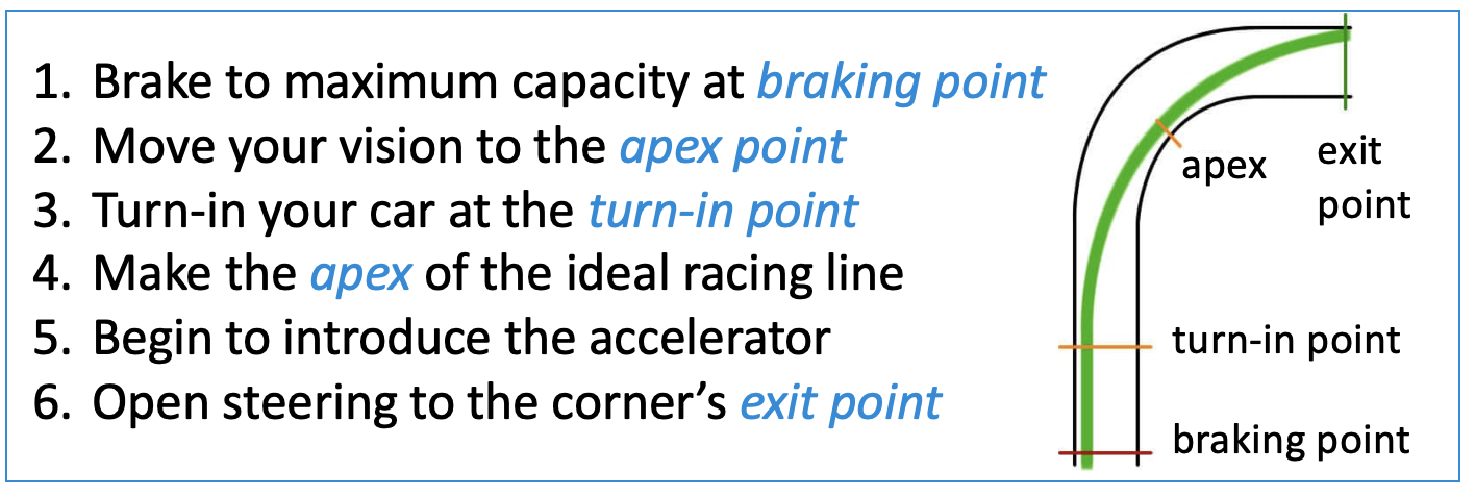}
    \caption{Sample of strategies taught at various online racing schools.}
    \label{fig:racing_strategies}
\end{figure}

A snippet of common racing strategies\footnote{As taught at online racing schools such as Driver 61: \href{https://driver61.com/uni/racing-line/}{\url{https://driver61.com/uni/racing-line/}}} taught to new racers is depicted in Fig.~\ref{fig:racing_strategies}. The most basic rules are guidelines for how to find the racing line, or the fastest possible path around the track, which is directly determined by the track curvature.
Drivers are instructed where along the track to brake, steer, and accelerate, and how to find these locations for curves of different shapes. 

The environmental constraints imposed on the vehicle (lane boundaries) are parameterized by the track curvature.
This environmental descriptor gives rise to physical areas along the track with respect to which racing rules are then explained, e.g. ``brake at the \textit{braking point}, then cut the curve at the \textit{apex} and aim for the outside of the straightaway." 
The strategies here consist of sections of the track towards which to aim the vehicle as well as acceleration profiles to apply along the way.
Importantly, the locations of these regions only depend on the local curvature; the track curvature a mile away has little impact on the location of the apex in the curve directly ahead.

We also note that the strategies are explained using only a subset of the state space: the distance from the centerline. 
Given guidelines on this subset, the driver is free to adjust other states and inputs such as vehicle velocity and steering in order to satisfy environmental constraints.

\subsection{Principles of Strategy}\label{ssec:approach}
Based on this real-life intuition, we propose three principles of navigation strategy:
\begin{enumerate}
    \item Strategies are a function of a local environment forecast (\textit{e.g. radius of curvature of an upcoming track segment})
    \item Strategies work in a reduced-order state space \\ (\textit{e.g. distance from center lane})
    \item Strategies provide target regions in the (reduced-order) state space for which to aim, and input guidelines for getting there \\ (\textit{e.g. ``braking point", ``turn-in point", ``exit point")} 
\end{enumerate}
The control architecture proposed in this paper formalizes the above principles of strategy, and shows how to incorporate such strategies into a hierarchical learning control framework.
In particular, we focus on two aspects. First, we show how to learn generalizable strategies from stored executions of previous control tasks.
Second, we show how to integrate the learned strategies in an MPC framework so as to guarantee feasibility when solving a new control task in real-time.

\subsection{Implementation}

Hierarchical Predictive Learning (HPL) is a data-driven control scheme based on applying high-level strategies learned from previous executions of different tasks. The HPL controller modifies its behavior whenever new strategies become applicable, and operates in coordination with a safety controller to ensure constraint satisfaction at all future time steps.
% \begin{figure}
%     \centering
%     \includegraphics[width=\columnwidth]{flowChart.png}
%     \caption{The Hierarchical Predictive Learning (HPL) control architecture. At time $k$, the state $x_k$ and $N$-step environment forecast ${{\theta}}_{k:k+N}$ are used to select a control strategy. 
% A strategy consists of reduced dimensions sets, $\tilde{\mathcal{X}}_{k+T}$ and $\tilde{\mathcal{U}}_{k:k+T}$, towards which to steer the system in the next $T$ time steps and input guidelines for getting there. 
% These sets are used to construct a full-dimension target set, ${\mathcal{X}}_{k+T}$, used as a terminal set in an MPC controller with horizon $N_{\mathrm{MPC}}$. At each time $k$, $\mathrm{SetList}_k$ determines the relationship between $T$ and $N_{\mathrm{MPC}}$.}
%     \label{fig:flow_draft}
% \end{figure}

\begin{figure}
    \centering
    \includegraphics[width=\columnwidth]{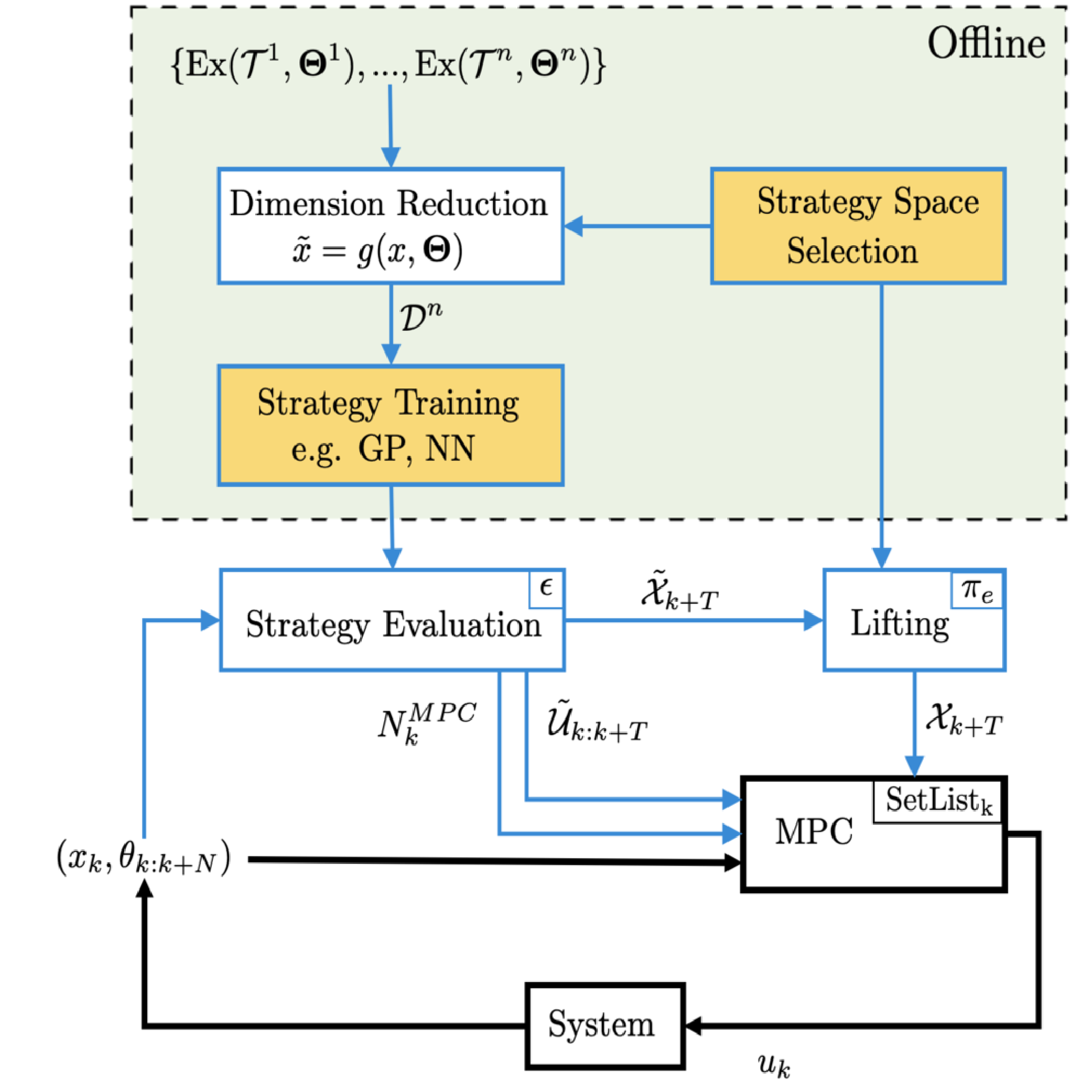}
    \caption{The Hierarchical Predictive Learning (HPL) control architecture. At time $k$, the state $x_k$ and $N$-step environment forecast ${{\theta}}_{k:k+N}$ are used to evaluate the control strategy. 
    A strategy consists of reduced dimensions sets, $\tilde{\mathcal{X}}_{k+T}$ and $\tilde{\mathcal{U}}_{k:k+T}$, towards which to steer the system in the next $T$ time steps and input guidelines for getting there. 
    These sets are used to construct a full-dimension target set, ${\mathcal{X}}_{k+T}$, used as a terminal set in an MPC controller with horizon $N^{\mathrm{MPC}}$. At each time $k$, $\mathrm{SetList}_k$ determines the relationship between $T$ and $N^{\mathrm{MPC}}$. 
    The low-level control loop is drawn in black, and shaded yellow blocks indicate control design choices.}
    \label{fig:flow_draft}
\end{figure}

An overview of the control architecture is shown in Fig.~\ref{fig:flow_draft}. 
Offline, before beginning the new control task, stored executions from previous tasks are used to train strategy functions of a desired parameterization (Sec.~\ref{sec:strategyfinder}). 
After training, the controller can solve new tasks.
Online, at each time $k$, an $N$-step local environment forecast ${{\theta}}_{k:k+N}$ (\ref{eq:env_forecast}) is used to determine if a new high-level control strategy is available (Sec.~\ref{sec:strategymanager}). 
A strategy consists of state and input sets in reduced dimensions, $\tilde{\mathcal{X}}_{k+T}$ and $\tilde{\mathcal{U}}_{k:k+T}$, that provide a set towards which to steer the system in the next $T$ timesteps, as well as input guidelines for getting there.
The strategy sets are used to construct a target set in the full state space, ${\mathcal{X}}_{k+T}$.
We note the difference between the environment forecast horizon $N$ and the strategy prediction horizon $T$ (Sec.~\ref{ssec:stratHor}). 
Lastly, an MPC controller with prediction horizon $N^{MPC}$ calculates a low-level input $u_k$ to reach the target set (Sec.~\ref{sec:lowlevelcontrol}).

There are five key design and control challenges:
\begin{enumerate}
    \item Representing the strategy mathematically
    \item Choosing appropriate horizons $N$, $T$, and $N^{MPC}$
    \item Lifting the reduced-dimension strategy set into a full-dimension target set
    \item Ensuring closed-loop strategy effectiveness at solving the new task
    \item Ensuring constraint satisfaction and recursive feasibility of the receding horizon control problem
\end{enumerate}
In the following sections, we address each of these challenges in detail and prove the feasibility of the resulting HPL control law. First, we show how to learn generalizable strategies from stored executions of previous tasks. Second, we show how to integrate the learned strategies in an MPC framework so as to guarantee feasibility when solving a new control task in real-time.

\section{Learning Strategies From Data}\label{sec:strategyfinder}

This section addresses the first aim of our paper: learning generalizable strategies from stored data of previously solved tasks. 
We consider strategies to be maps from a state and environment forecast to reduced-dimension strategy sets:
\begin{align}\label{eq:strategyform}
    (\tilde{\mathcal{X}}_{k+T}, ~\tilde{\mathcal{U}}_{k:k+T}) = \mathcal{S}(x_k,~\theta_{k:k+N}),
\end{align}
where $x_k$ is the system state at time $k$ and $\theta_{k:k+N}$ the $N$-step environment forecast. $\tilde{\mathcal{X}}_{k+T}$ represents a strategy set in reduced dimension that the system should be in $T$ timesteps into the future, and $\tilde{\mathcal{U}}_{k:k+T}$ provides constraint guidelines for a reduced dimension of the input as the system travels towards $\tilde{\mathcal{X}}_{k+T}$.

There are several ways of representing the strategy function $\mathcal{S}$ in  (\ref{eq:strategyform}), including model-based methods that use an explicit model for how variations in task environments affect the optimal control input \cite{bertsimas2018voice}.
In this work we instead opt for a data-driven approach, using stored executions (\ref{eq:iterationvecs}) that solve related tasks. 

\subsection{Strategy States and Inputs}
The strategy function outputs reduced-dimension state and input sets. We refer to the space where these reduced-dimensional sets lie as ``strategy state space" and ``strategy input space." 

We define the strategy state at each time $k$ as
\begin{align}\label{eq:stateprojection}
    \tilde{x}_k = g(x_k,{\bf{\Theta}}) \in \mathbb{R}^{n_{\tilde{x}}},
\end{align}
where $g$ maps the full-dimensional state $x_k$ into the corresponding lower-dimensional strategy state $\tilde{x}_k$. Similarly, the strategy inputs are
\begin{align}\label{eq:inputprojection}
    \tilde{u}_k = r(u_k, {\bf{\Theta}}) \in \mathbb{R}^{n_{\tilde{u}}},
\end{align}
where $r$ maps the full-dimensional input at time $k$ into lower-dimensional strategy inputs.
These mapping functions may depend on the task's environment descriptor function ${\bf{\Theta}}$ - for example, a strategy state $\tilde{x}$ measuring the distance from a race track centerline depends on the shape of the centerline, which is described by ${\bf{\Theta}}$.
We denote the strategy state and strategy input spaces as
\begin{align}\label{eq:sxset}
    \tilde{\mathcal{X}} &= \{\tilde{x} ~| ~\tilde{x} = g(x, {\bf{\Theta}}),~ x \in \mathcal{X} \},\\ 
    \tilde{\mathcal{U}} &= \{\tilde{u} ~|~ \tilde{u} = r(u, {\bf{\Theta}}),~ u \in \mathcal{U} \}.
\end{align}

\begin{remark}
The functions $g$ and $r$ may be any functions mapping $\mathcal{X}$ and $\mathcal{U}$ to a reduced-dimension space. They are not limited to indexing functions, i.e. we do not require $\tilde{x}_k$ to be a subset of $\mathcal{X}$.
\end{remark} 

We use strategy states and inputs (rather than the full states and inputs directly) to allow for simplification and generalization. While all system states and inputs affect the system's trajectory via the dynamics (\ref{eq:VehicleModel}), when solving complex tasks a subset of states and inputs are likely to be especially informative as to how the system should respond to the upcoming environment. 

The choice of strategy states and inputs is therefore critical. Strategy states and inputs must be meaningful to solving the task at hand, and it must be reasonably expected that a strategy mapping (\ref{eq:strategyform}) from an environment forecast to these strategy spaces can be formulated.
If the chosen strategy states and inputs are not correlated with (i.e. can not be predicted from) the environment forecast, new strategy states must be chosen.

%Choosing strategy states is one possibility for using human intuition about the specific task to inform the control design. 

Initially, all system states and environment descriptions can be considered as candidate strategy states.  
Because reducing the number of strategy states can improve generalizability, this list should then be trimmed as much as possible (or as much as required for computational tractability). 
Human knowledge can provide insight into what states are likely to have an impact on optimal behavior, and therefore which strategy states and inputs to choose.
In the autonomous racing task, for example, a control designer will know that it is easier to use the track curvature forecast to predict the vehicle's future distance along and from the centerline than a future yaw rate.
Note that strategy states should encode information about both task objective and task safety.

If no human intuition can be incorporated, data-driven methods could instead be used to find appropriate strategy states and inputs. 
Clustering methods or dimensionality-reduction methods such as PCA can estimate what states are most important for determining the strategy.
Recent work \cite{bertsekasagg}, for example, proposes forming aggregate (or representative) features out of system states in order to reduce the problem dimension in Dynamic Programming - this approach could similarly be applied to finding strategy states for HPL.
Using data-driven methods to determine the best choice of strategy states, particularly for complex tasks, is an avenue for future research. 

In general, we suggest using intuition about the task to narrow the list of potential strategy states as much as possible, and using data-driven feature selection methods in the final stages if strategy performance using only hand-derived features is lacking. 
Sections \ref{sec:robot}-\ref{sec:flappybird} provide several examples for choosing strategy states for navigation tasks.

\begin{remark} It may also be beneficial to use the current strategy state $\tilde{x}_k$, rather than the current state $x_k$, as an input to the strategy function (\ref{eq:strategyform}). This can further improve the generalizability of the learned strategy, as fewer values have to match between the new environment and the stored task data (\ref{eq:iterationvecs}). 
The notation in the remainder of this paper constructs the strategy input using the full state $x_k$, but the approach and theory remain the same if the strategy state space is used instead.
\end{remark}

\subsection{Representing the Strategy}
Thanks to an overwhelming amount of machine learning research, there are myriad ways to represent the strategy function $\mathcal{S}$ in (\ref{eq:strategyform}).
We propose using Gaussian Processes (GPs).
GPs have frequently been used in recent predictive control literature to provide data-driven estimates of unknown nonlinear dynamics \cite{hewing2019cautious, learningGaussProcess,klenske2015gaussian}. 
%\textcolor{red}{Mention the discussion about being able to correctly learn models with some probability, cite the papers - to include the eqn?}
Specifically, GPs are used to approximate vector-valued functions with real (scalar) outputs.

Given training data (input vectors and output values), GPs use a similarity measure known as ``kernel" between pairs of inputs to learn a nonlinear approximation of the true underlying input-output mapping.
The kernel represents the learned covariance between two function evaluations, and is parameterized by a set of hyperparameters. 
Once the hyperparameters of the kernel have been optimized, the GP can be queried at a new input vector. 

When evaluated at a new input, the GP returns a Gaussian distribution over output estimates; thus GPs provide a best guess for the output value corresponding to an input (mean) and a measure of uncertainty about the estimate (variance).
This allows us to gauge how confident the GP is in its prediction at a particular input, a critical component of the HPL framework. A review of GPs in control is provided in \cite{kocijan2016modelling}.

We note that the robust MPC community often prefers stochastic models with bounded support \cite{bujarbaruahAdapFIR} to GPs, in order to have strict safety guarantees. Our approach can be extended to these types of models as well. 

% The GPs are trained \textit{offline}, and the strategy sets are built \textit{online} during the execution of a new control task using the estimated mean and standard deviation provided by the GPs. 

% How to build the training data
\subsection{Training the Strategy (Offline)}\label{ssec:trainingGP}

The GPs are trained to predict the values of the strategy states (\ref{eq:stateprojection}) and inputs (\ref{eq:inputprojection}) at $T$ timesteps into the future, based on the current state and $N$-step environment forecast.
Each GP approximates the mapping to one strategy state or input:
\begin{align}\label{eq:gpeq}
    (\mu, \sigma^2) = \mathrm{GP}(x_k, \theta^i_{k:k+N}),
\end{align}
where $\mu$ and $\sigma^2$ represent statistics of the Gaussian distribution over strategy state estimates.
GPs best approximate functions with scalar outputs, so we train one GP for each strategy state and input (a total of $n_{\tilde{x}} + n_{\tilde{u}}$ number of GPs). 
The learned GPs capture high-level strategies that were common to a variety of previously solved, related tasks. 
We note that the GPs are evaluated only at the current state and environment forecast - they are time-invariant and do not depend on the new task beyond the environment forecast. 

We use the stored executions (\ref{eq:iterationvecs}) from previous control tasks to create GP training data. The training output data for each GP contains the strategy state or input the GP is learning to predict. 
After solving $n$ control tasks (see Remark 1), the training data consists of: 
\begin{align}\label{eq:dataset}
    \mathcal{D} = \{{\bf{z}} & = [z^1_0, z^1_1, \dots,  z_{D^1-T}^1, z_0^2, \dots, z_{D^n-T}^n]^\top \\ 
    {\bf{y}} & = [y^1_0, y^1_1, \dots, y_{D^n-T}^n]^\top. \nonumber
\end{align}
Each input vector $z^i_k$ corresponds to the output $y^i_k$, where
\begin{align}\label{eq:datainput}
    z^i_k = [(x^i_k)^\top, (\theta^i_k)^\top, (\theta^i_{k+1})^\top, ..., &(\theta^i_{k+N})^\top], \\
    & ~ i \in [1,n], ~ k \in [0, D^i - T],\nonumber
\end{align}
and $\theta^i_k$ denotes the local environment at time $k$ of the $i$th control task. We note that in (\ref{eq:datainput}), each GP uses the same training input data, but this need not be the case. 

The corresponding output entry $y^i_k$ contains the value of the strategy state of interest at $T$ time steps in the future:
\begin{align}
    y^i_j = (\tilde{x}^{i}_{k+T})^\top,  ~ i \in [1,n], ~ k \in [0, D^i - T].\nonumber
\end{align}
The input strategy set can be similarly parameterized in a number of different ways, such as minimum and maximum values realized over the $T$-step trajectory. 

Once the training data (\ref{eq:datainput}) is collected, the GP kernel hyperparameters are optimized using maximum log-likelihood regression. 
In this paper, we use the squared-exponential kernel, though different kernels can be chosen depending on the expected form of the task-specific strategy equation (\ref{eq:strategyform}). 
Given two entries of {\bf{z}} in (\ref{eq:dataset}), the squared-exponential kernel evaluates as
\begin{align}
    k(z^o_i, z^w_j) = \sigma_f^2 \exp{-\frac{1}{2} \sum_{m=1}^{n_x+N+1} \frac{(z^o_{i}(m) - z^w_{j}(m))^2}{\sigma_m^2}},\nonumber
\end{align}
where $z^o_{i}(m)$ is the $m$th entry of the vector $z^o_i$.
Many software packages exist that automate Bayesian optimization of the hyperparameters, including the Machine Learning Toolbox in Matlab and SciKit Learn or GPyTorch in Python. 

\subsection{Evaluating the Strategy (Online)} \label{ssec:stratsets}
Once trained, the GPs can be evaluated on data from a new task. 
At time $k$ of a new task $\mathcal{T}^{n+1}$, we evaluate the GPs at the new query vector $z^{n+1}_k$, formed as in (\ref{eq:datainput}), to construct hyperrectangular strategy sets in reduced-dimension space.

\begin{figure}
    \centering
    \includegraphics[width = 0.9\columnwidth]{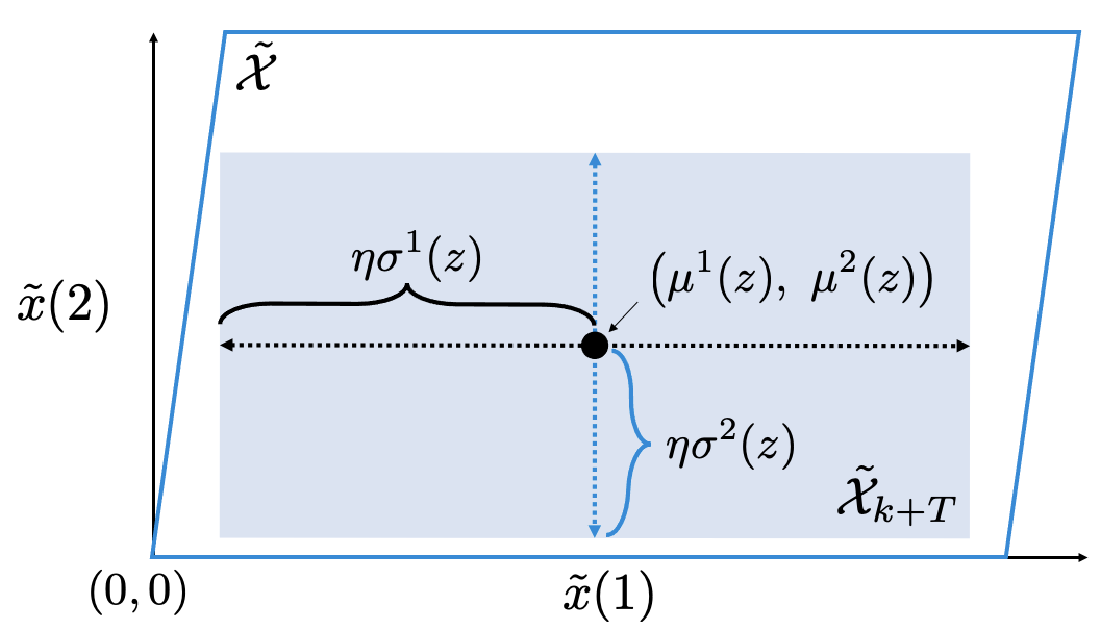}
    \caption{Each dimension $\tilde{x}(i)$ of the strategy set $\tilde{\mathcal{X}}_{k+T}$ is bounded using the mean and variance of a GP evaluation (\ref{eq:strategystateconstraint}).}
    \label{fig:hyperrect}
\end{figure}

Each GP returns a one-dimensional Gaussian distribution over output scalars, parameterized by a mean $\mu$ and variance $\sigma^2$.
Specifically, these are evaluated as:
\begin{align}\label{eq:gpeval}
    \mu(z^{n+1}_k) & = {\bf{k}}(z^{n+1}_k)\bar{K}^{-1}{\bf{y}} \\
    \sigma(z^{n+1}_k)^2 &= k(z^{n+1}_k,z^{n+1}_k) - {\bf{k}}(z^{n+1}_k)\bar{K}^{-1}{\bf{k}}^\top(z^{n+1}_k),
\end{align}
where 
\begin{align}
    {\bf{k}}(z^{n+1}_k) = [k(z^{n+1}_k, z^1_0), \dots, k(z^{n+1}_k, z^n_{D^n-T})],
\end{align}
and the matrix $\bar{K}$ is formed out of the covariances between training data samples such that
\begin{align}
    \bar{K}_{i,j} = k({\bf{z}}_i, {\bf{z}}_j). 
\end{align}
Given means and variances, we form one-dimensional bounds on each $i$th strategy state and $j$th strategy input as
\begin{align}
    \tilde{\mathcal{X}}_{k+T}(i) &= [\mu^i(z_k^{n+1}) \pm \eta \sigma^i(z_k^{n+1})],~  \forall i \in [1, n_{\tilde{x}}] \label{eq:strategystateconstraint}\\
    \tilde{\mathcal{U}}_{k:k+T}(j) &= [\mu^{j}(z_k^{n+1}) \pm \eta \sigma^{j}(z_k^{n+1})],~  \nonumber \\
    & ~~~~~~~~~~~~~~~~~~~~~~~~~~~\forall j \in [n_{\tilde{x}}+1, n_{\tilde{x}} + n_{\tilde{u}}]. \nonumber
\end{align}
In (\ref{eq:strategystateconstraint}), $\mu^i(z_k^{n+1})$ and $\sigma^i(z_k^{n+1})$ are the means and standard deviations computed by the $i$th GP evaluated at $z_k^{n+1}$. 
The parameter $\eta>0$ determines the size of the range.
When these one-dimensional bounds are combined for all strategy states and strategy inputs, hyperrectangular strategy sets are formed in strategy space, with each dimension constrained according to (\ref{eq:strategystateconstraint}).
An example is shown in Fig.~\ref{fig:hyperrect}.
The hyperrectangular strategy sets are denoted $\tilde{\mathcal{X}}_{k+T}$ and $\tilde{\mathcal{U}}_{k:k+T}$, and indicate where (in strategy space) the system should be in $T$ timesteps and what inputs to apply to get there.

\subsection{Other Strategy Models} \label{sec:strategy_models}
There are many benefits to using GPs to model the strategy function (\ref{eq:strategyform}), including the interpretability of the output variance as a level of strategy confidence. 
One downside to using GPs is that they are very computationally inefficient to train: the hyperparameter optimization time scales as the cube of the training data size. 
This forces a trade-off between computational complexity and the effectiveness of the learned strategy, which benefits from seeing as much training data as possible. 
It also prevents easy online fine-tuning of the learned strategy as more task data becomes available.

If training complexity becomes too cumbersome, the strategy function (\ref{eq:strategyform}) may be replaced by a different data-driven model that best fits the requirements of the considered task. In addition to computational complexity, there are two main considerations for choosing a strategy model for HPL: 
\begin{enumerate}
    \item the model should be able to approximate our best reasonable guess for the true shape of the underlying function (e.g. bounded, continuous, etc.), and
    \item the model should have a way of estimating the confidence in a particular strategy evaluation.
\end{enumerate}
Any data-driven model that appropriately satisfies these two requirements may be used in place of the GP without affecting the theory presented in this paper.

\subsubsection*{Neural Networks}
With regard to the first consideration, using universal function approximators such as neural networks is easy and appealing, as they can be composed to represent functions of most shapes (at the cost of more hyperparameters to optimize). 
However, estimating the confidence associated with a particular strategy prediction is much more difficult for neural networks than for GPs \cite{nandeshwar2006models}. 
Common approaches include bootstrapping \cite{ott2018analyzing} or training additional network layers to estimate the confidence \cite{extraLayers1}.
These approaches are often tuned to the considered task, and can be poorly calibrated \cite{papadopoulos2001confidence}.
Various efforts to improve confidence estimation in neural networks have been made, but these methods often increase the size of the learned network, which worsens computational complexity \cite{towardsBetter, cortes2018deep}. 
A promising new avenue is recent work in model-based neural networks \cite{robey2020model}, but further validation is required.

\subsection{Strategy Horizons}\label{ssec:stratHor}
The strategy mapping uses two different horizons: the environment forecast horizon $N$ and the strategy prediction horizon $T$. 
$N$ determines how much information about the future environment the strategy takes into account, while $T$ determines how far out the strategy predicts the values of strategy states and inputs. The two need not be the same. In fact, for many navigation tasks it makes sense to choose $N > T$. 

In the autonomous racing example, $N>T$ corresponds to looking further ahead along the track curvature and then only planning a trajectory along the first part of the visible track. This choice means that the strategy can suggest a strategy set to aim for that is also likely to be feasible in and optimal for the immediate future, since the upcoming environmental constraints were already implicitly considered.
Choosing $N=T$, in contrast, means that the strategy needs to predict a strategy set as far out as the environment is known. If the environment  changes immediately beyond the horizon $N$, the proposed strategy set $\tilde{X}_{k+T}$ may have been a poor choice.

As with choosing strategy states, human intuition can also play a role in choosing appropriate strategy horizons. Alternatively, the collected executions (\ref{eq:iterationvecs}) from previous tasks can be examined to determine appropriate values for $N$ and $T$, by evaluating how great environmental differences must be before the system's trajectory changes. This can typically be estimated from the available task data, and fine-tuned as necessary using cross-validation.
In general, it is wise to use the largest computationally-allowable $N$.

Sections \ref{sec:robot}-\ref{sec:flappybird} provide several examples for choosing strategy horizons.

\begin{remark}
As described thus far, both horizons $N$ and $T$ are time horizons, i.e. they forecast the environment and predict the strategy state at certain numbers of time steps in the future.
It is also possible to, instead, forecast the environment and strategy state at a fixed distance into the future, or use a mix of time- and space-forecasting for $N$ and $T$.
In the racing example, this could correspond to always seeing a fixed number of meters ahead, no matter the vehicle speed.
This approach is considered in Sec.~\ref{sec:form1}. 
\end{remark}

\section{Safely Applying Learned Strategies}\label{sec:strategymanager}

We now address the second aim of our paper: using the strategy sets in a low-level controller while maintaining safety guarantees. 
Our approach consists of \textit{i)} lifting the reduced-dimension strategy sets (\ref{eq:strategystateconstraint}) back into the full-dimensional state space, and \textit{ii)} integrating the lifted strategy set with a safety controller. The result is a target set that can be used in a low-level MPC controller.

%Our method relies on the existence of a safety controller and corresponding safe set.
\begin{assumption}\label{ass:safetycontroller}
There exists a safety control policy that can prevent the system (\ref{eq:VehicleModel}) from violating both system- and task-specific environment constraints (\ref{eq:taskconstraints}). 
In particular, there exists a safe set 
\begin{equation}\label{eq:emergencyset}
    \mathcal{X}_E \subseteq \mathcal{X}({\bf{\Theta}}),
\end{equation}
and a corresponding safety control policy
\begin{equation}\label{eq:emergencycontrol}
    u = \pi_{\mathrm{e}}(x, {\bf{\Theta}}),
\end{equation}
such that $\forall x \in \mathcal{X}_E,~ f(x, \pi_{\mathrm{e}}(x, {\bf{\Theta}})) \in \mathcal{X}_E.$
% \begin{align}\nonumber
%     \forall x \in \mathcal{X}_E,~ f(x, \pi_{\mathrm{e}}(x, \theta)) \in \mathcal{X}_E.
% \end{align}
\end{assumption}
% The safe set and safety control law are system- and task-specific. They may be parameterized by the environment descriptor $\theta$, but the safety control law (\ref{eq:emergencycontrol}) and safety set (\ref{eq:emergencyset}) must be the same for all possible instantiations of $\theta$-parameterized environments.
% For example, a safety control law may bring the system (\ref{eq:VehicleModel}) to a physical stop before continuing to complete a suboptimal task execution. 

\begin{remark}
Given a safety controller (\ref{eq:emergencycontrol}), a safe set (\ref{eq:emergencyset}) may be found using a variety of data-driven methods, such as sample-based forward reachability from a gridded state space $\mathcal{X}$ \cite{data_invariant_sets, wang2020datadriven, ozay2018}.
%. The reachability analysis can be repeated until $\mathcal{X}_E$ converges, or for a maximum number of iterations.
%Another possibility is to grid the state space $\mathcal{X}$ and use forward reachability to find states which lead to feasible executions (\ref{eq:iterationvecs}) using (\ref{eq:emergencycontrol}). 
%The set $\mathcal{X}_E$ is formed out of the gridded state which lead to closed-loop trajectories with (\ref{eq:emergencycontrol}) that satisfy state and input constraints (\ref{eq:VehicleModelConstr},~\ref{eq:taskconstraints}). 
\end{remark}

\subsection{Lifting Strategy Sets to Full-Dimensional Target Sets} 
\begin{figure}
    \centering
    \includegraphics[width=0.93\columnwidth]{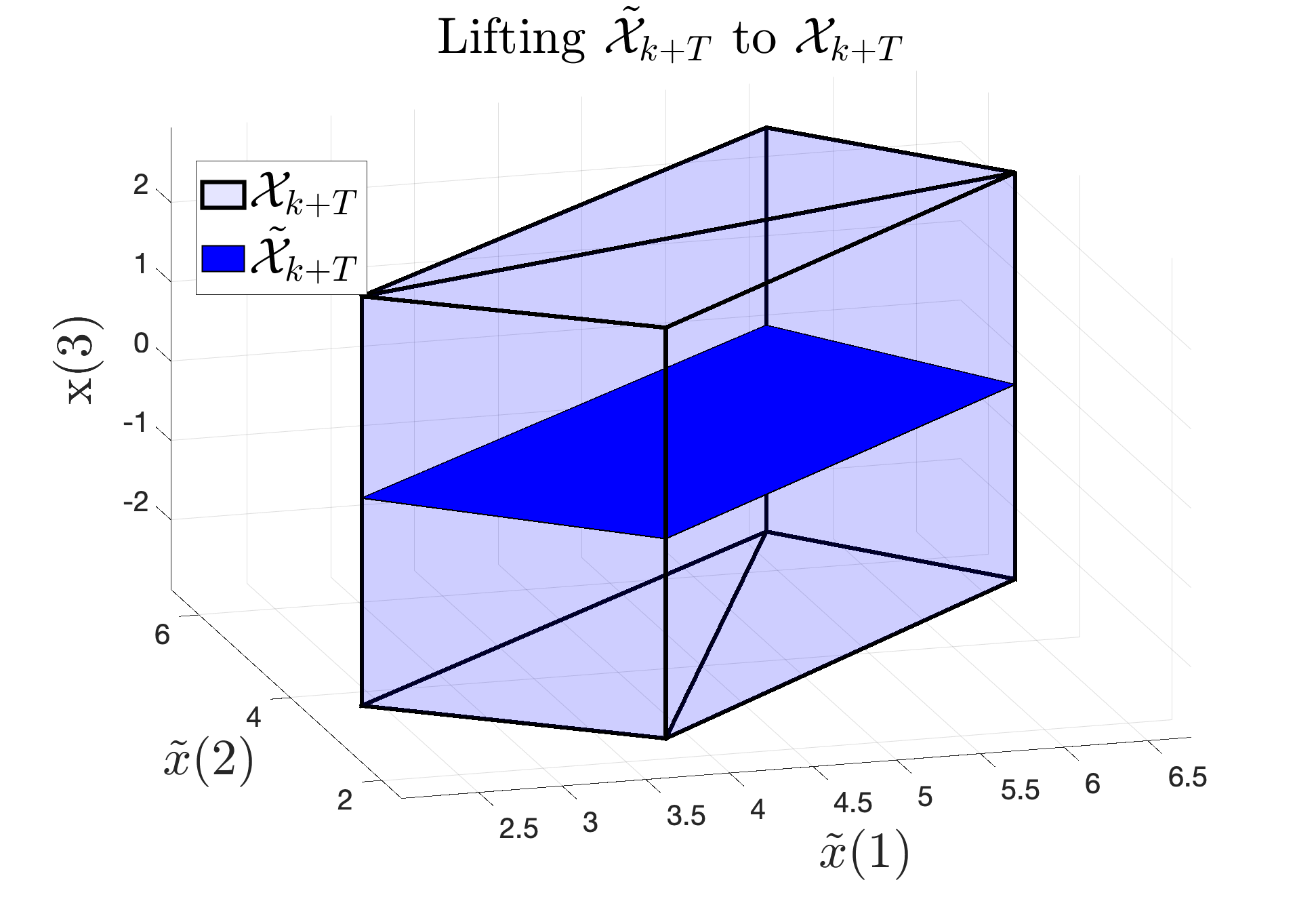}
    \caption{In the lifted strategy set (\ref{eq:xt}), the strategy states $\tilde{x}(1)$ and $\tilde{x}(2)$ are constrained to lie in the strategy set, with additional states like $x(3)$ constrained according to $\mathcal{X}_E$.}
    \label{fig:3dvxplot}
\end{figure}

At each time $k$ of solving a task $\mathcal{T}^{n+1}$, new reduced-dimension strategy sets are constructed according to (\ref{eq:strategystateconstraint}). 
These strategy sets must be lifted to target sets in the full-dimensional state space so they can be used as a terminal constraint in a low-level MPC controller. 
%Critically, this target set must also be \textit{safe}: in order to guarantee feasibility of our controller for all future timesteps, we require that the target set is a subset of the safe set (\ref{eq:emergencyset}). 
% The target set has two properties:
% \begin{enumerate}
%     \item The target set belongs to the safe set (\ref{eq:emergencyset}).
%     \item It must be reachable by the system (\ref{eq:VehicleModel}) in $T$ time steps.
% \end{enumerate}
Critically, the target set must belong to the safe set (\ref{eq:emergencyset}). This ensures that once the system has reached the target set, there will always exist at least one feasible input (the safety control (\ref{eq:emergencycontrol})) that allows the system to satisfy all state constraints.
Given strategy sets (\ref{eq:strategystateconstraint}), we find a corresponding lifted strategy set:
\begin{align}\label{eq:xt}
    \mathcal{X}_{k+T} = \{x \in \mathcal{X}_E ~| ~g(x, {\bf{\Theta}}) \in \tilde{\mathcal{X}}_{k+T} \},
\end{align}
where $g(x, {\bf{\Theta}})$ is the projection of the full-dimensional state $x$ onto the set of chosen strategy states, as in (\ref{eq:stateprojection}). 
$\mathcal{X}_{k+T}$ is a full-dimensional set in which the strategy states (\ref{eq:stateprojection}) lie in the GP's strategy sets (\ref{eq:strategystateconstraint}) and the remaining states are in the safety set. 
Thus for any state $x_k \in \mathcal{X}_{k+T}$, the safety control (\ref{eq:emergencycontrol}) can be applied if necessary to ensure constraint satisfaction in future time steps. 
Figure \ref{fig:3dvxplot} depicts the difference between a strategy set and its lifted strategy set. 

\subsection{Tuning Risk}\label{ssec:betarisk}

If desired, control designers can specify a maximum risk level $\beta \in [0,1]$ to control how conservative the target set $\mathcal{X}_{k+T}$ is with regards to incorporating the safety control. The above formulation (\ref{eq:xt}) for the terminal set corresponds to \textit{no} risk (i.e. $\beta = 0$), since for all states in $\mathcal{X}_{k+T}$ there exists at least one feasible input sequence (the safety control) that will result in a safe closed-loop state evolution. 
%However, this conservative formulation may be too restrictive. 

If this is too restrictive, the target set can be chosen to be a convex combination of the safety set $\mathcal{X}_E$ and the environmental state constraint set: 
\begin{align}\label{eq:betaxt}
    \mathcal{X}_{k+T} = \{x \in \beta\mathcal{X}_E + (1-\beta)\mathcal{X}({\bf{\Theta}})& ~| ~ \\
    & g(x,{\bf{\Theta}}) \in \tilde{\mathcal{X}}_{k+T} \} \nonumber,
\end{align}
By varying $\beta$ in the range between 0 and 1, we vary how conservative our approach needs to be, depending on the cost of task failure.
As the value of $\beta$ increases, the target set converges to the state constraints imposed on the system by the task environment ($\beta = 1$). 

\begin{figure}
    \centering
    \includegraphics[width=0.9\columnwidth]{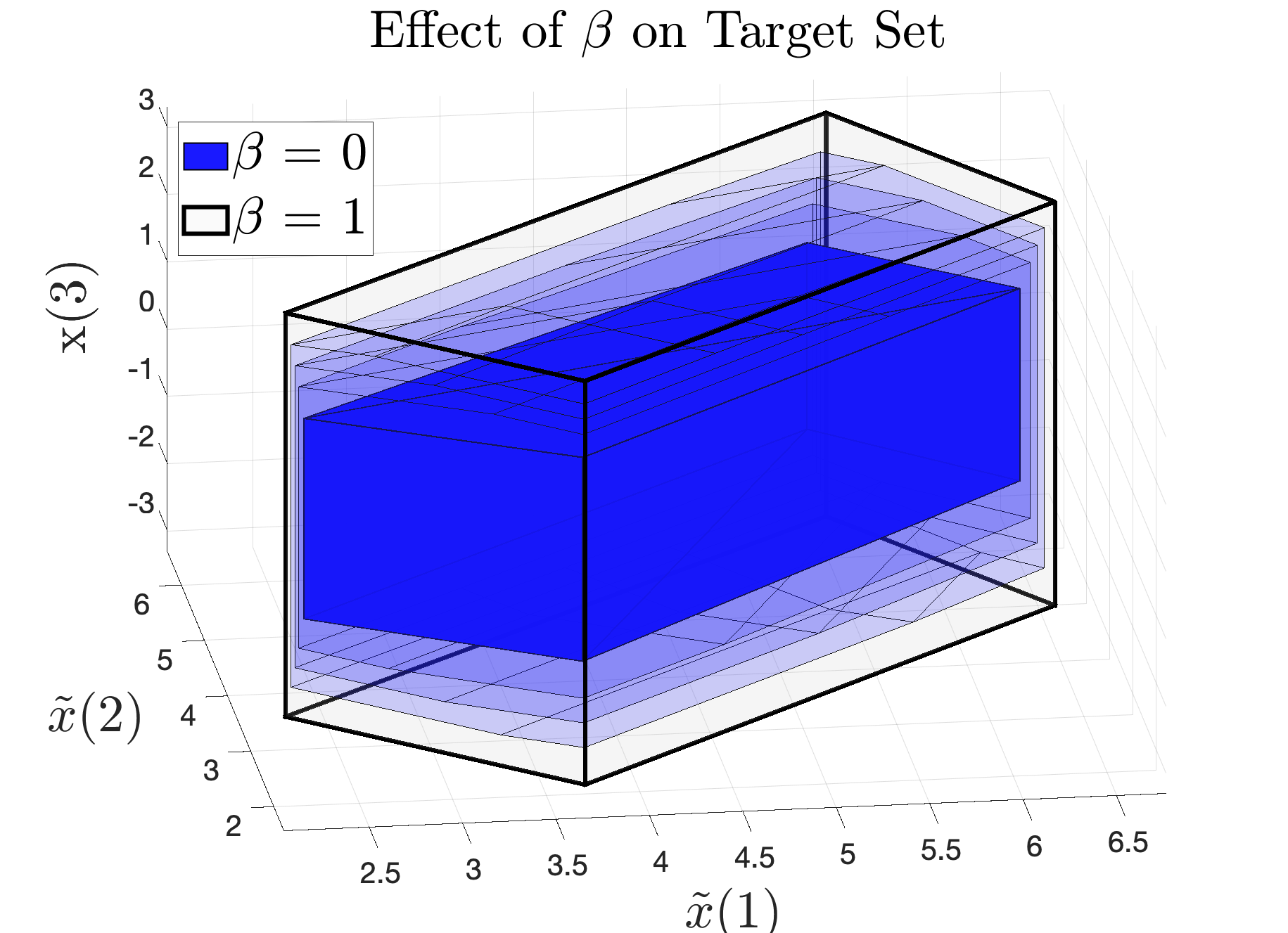}
    \caption{As $\beta$ varies, the constraints on the non-strategy state $x(3)$ are imposed either through $\mathcal{X}_E$ (if $\beta = 0$), $\mathcal{X}(\theta)$ (if $\beta = 1)$, or a combination of both (if $0 < \beta < 1$).}
    \label{fig:3dvelplot}
\end{figure}

An example of this interpolation is shown in Fig.~\ref{fig:3dvelplot}, which depicts how the constraints for the third entry of the system state $x(3)$ vary across different points in the strategy set (in the $\tilde{x}(1)-\tilde{x}(2)$ plane) and as $\beta$ varies. 
We see that as $\beta$ ranges from $0$ to $1$ the size of the non-strategy state constraint sets increases. When $\beta = 0$, plotted in dark blue, the target set is as in Fig.~\ref{fig:3dvxplot}. 

\subsection{Incorporating the Uncertainty Measure}
A benefit of using GPs to represent the strategy is that the standard deviation around an estimate may be used to evaluate how confident the GP is in its prediction at a particular input.
At time $k$, consider a vector $\mathcal{C}_k$ containing the standard deviations of the evaluated GPs:
\begin{align}\label{eq:uncertaintymeasure}
    \mathcal{C}_k = [\sigma^1(z_k^{n+1}), \dots , \sigma^{n_{sx}+n_{su}}(z_k^{n+1})]. 
\end{align}
If the GPs return a strategy set with standard deviations larger than a chosen threshold $d_{\mathrm{thresh}}$, we may opt not to use this strategy. 
% If the GPs return a strategy set estimate with standard deviations below a chosen thresholds $d_{\mathrm{thresh}}$, the Strategy Finder is sufficiently confident in the prediction. 
% We expect this to occur whenever the system has seen similar local environments in previous tasks and has learned a consistent strategy from those task executions.
% In this case the estimated LDS set is likely to be useful to the controller once it has been converted to a full-dimensional constraint set. 
% On the other hand, the strategy may have uncertainty measures above $d_{\mathrm{thresh}}$. 
We expect $\mathcal{C}_k > d_{\mathrm{thresh}}$ if either
\begin{enumerate}
    \item the system did not encounter a similar environment forecast in a previous control task (training data), or
    \item in previous control tasks this environment forecast did not lead to a single coherent strategy, resulting in a wide distribution of potential future strategy states.
\end{enumerate}
With high uncertainty measures, the strategy sets are not likely to contain valuable control information for the system. 
In this case, the target set is set to be \textit{empty}: $\mathcal{X}_{k+T} = [~]$.
The next section~\ref{sec:lowlevelcontrol} explains that this results in a horizon shift for the low-level MPC, and the system (\ref{eq:VehicleModel}) re-uses the target set from the previous time step.

%%%%%%%%%%%%%%%%%%%%%%%%%%%%%%%%%%%%%%%%%%%%%%%%%%%%%%%%%%
\section{Low-level Controller Design}\label{sec:lowlevelcontrol}
% Important: shifting terminal sets! We don't know if we will have a new strategy at each time step (depends on the confidence bound of the strategy).
% Shrinking horizons are a common strategy in MPC literature to deal with this. Works in the absence of model uncertainty; we had a solution at the previous time step so we simply re-use the same set and shorten the horizon. 
% We use the full-dimensional hyperrectangle as a terminal set; we choose our MPC horizon according to how far out we have a strategy for. 
% If we haven't found any new strategies in the last $T$ time steps (our MPC horizon is 0), we use the safety controller. We prove in the feasibility guarantees section that whenever this occurs, we are in the safe set. 

The low-level MPC controller calculates the input to be applied to the system at each time $k$.
This input is calculated based on the sequence of target sets (\ref{eq:xt}) found during the last $T$ timesteps.
%These sets inform where in the state space the system should be at each of the upcoming $T$ timesteps.

\subsection{Target Set List}
At each time $k$ of solving task $\mathcal{T}^{n+1}$, a new target set (\ref{eq:xt}) is constructed by lifting the strategy sets (\ref{eq:strategystateconstraint}). 
However, if the standard deviations (\ref{eq:uncertaintymeasure}) are too high, or there is no feasible input sequence to reach the target set $\mathcal{X}_{k+T}$, the target set for time $k$ will be empty: $\mathcal{X}_{k+T} = [~]$.

The target set list keeps track of the target sets (empty or not) which were constructed during the most recent $T$ timesteps:
\begin{align}\label{eq:setlist}
    \mathrm{SetList}_k = [{\mathcal{X}}_{k+1}, {\mathcal{X}}_{k+2}, \dots, {\mathcal{X}}_{k+T}].
\end{align}
At each new time step, the first set is removed and the target set found at the current time step $k$ is appended to the end. 
In this way, the target set list (\ref{eq:setlist}) always maintains exactly $T$ sets, though some (including the last set $\mathcal{X}_{k+T}$) may be empty.
This list is used to guide the objective function and constraints of the MPC controller. 

\subsection{Shifting Horizon MPC Formulation}
We formulate an MPC controller to calculate our input at each time step:
\begin{align}\label{eq:lowlevelmpc}
{\bf{u}}^\star(x_k) &= \argmin_{u_{0|k},...,u_{N^{MPC}_k-1|k}} \sum_{j\in \mathcal{S}_k}^{N^{MPC}_k-1}\mathrm{dist}\big(x_{j|k}, \mathcal{X}_{k+j}\big) \nonumber\\
& ~~~~~~ \mathrm{s.t.}~~ x_{j+1|k} = f(x_{j|k}, u_{j|k}) \nonumber\\ & ~~~~~~~~~~~~~~~ x_{j|k} \in \mathcal{X}({\bf{\Theta}}^{n+1}) ~~\forall j \in \{ 0, N^{MPC}_k-1\} \nonumber\\
& ~~~~~~~~~~~~~~~ u_{j|k} \in \mathcal{U}~~~~~~~~~~~ \forall j \in \{ 0, N^{MPC}_{k}-1\}\nonumber\\
& ~~~~~~~~~~~~~~~ x_{N^{MPC}_k|k} \in \mathcal{X}_{k+N^{MPC}_k}\nonumber \\
& ~~~~~~~~~~~~~~~ x_{0|k} = x_k,\
\end{align}
where $\mathcal{S}_k$ is the set of indices with non-empty target sets, 
\begin{align}
    \mathcal{S}_k = \{s~ |~ \mathrm{notEmpty}(\mathcal{X}_{k+s-1}) \}. \nonumber
\end{align}
The MPC objective function (\ref{eq:lowlevelmpc}) penalizes the Euclidean distance from each predicted state to the target set corresponding to that prediction time. For a smoother cost, the objective could be augmented to take the input effort into account.

The MPC uses a time-varying shifting horizon $0 < N^{MPC}_k \leq T$ that corresponds to the largest time step into the future for which a non-empty target set results in feasibility of (\ref{eq:lowlevelmpc}):
\begin{align}\label{eq:nk}
    {N^{MPC}_k} = \max s~:~ \{s \in \mathcal{S}_k,~ (\ref{eq:lowlevelmpc}) \text{ is feasible with $N^{MPC}_k = s$} \}. 
\end{align}
This ensures that the MPC controller (\ref{eq:lowlevelmpc}) has a non-empty terminal constraint and the optimization problem is feasible. 
To avoid unnecessary repeated computations, all target sets in the target set list (\ref{eq:setlist}) which lead to infeasibility of (\ref{eq:lowlevelmpc}) when used as the terminal constraint are set as \textit{empty} in (\ref{eq:setlist}).
At time step $k$, we apply the first optimal input to the system:
\begin{align}\label{eq:lowlevelinput}
    u_k = u^\star_{0|k}.
\end{align}

\begin{remark}
Here, new target sets (\ref{eq:xt}) are found at the same frequency as the controller update (\ref{eq:lowlevelmpc})-(\ref{eq:lowlevelinput}), but this can easily be adapted for asynchronous loops as in \cite{ugo2020unified}.
\end{remark} 

\begin{remark}
Evaluating GPs to calculate the strategy is fast when optimized, and constructing strategy sets according to (\ref{eq:xt}) is straightforward once GPs have been evaluated. Incorporating strategies will thus not have significant impacts on the closed-loop run time of the MPC controller (\ref{eq:lowlevelmpc})-(\ref{eq:lowlevelinput}).
\end{remark}

\subsection{Safety Control}
%The MPC horizon is adjusted whenever no feasible, high-confidence target set (\ref{eq:xt}) can be constructed. 
If no target sets in (\ref{eq:setlist}) can feasibly be used as a terminal constraint in (\ref{eq:lowlevelmpc}), all sets in the target set list will be empty, and the MPC horizon is $N^{MPC}_k=0$.
When this occurs, the system enters into \textit{Safety Control mode}. The safety controller (\ref{eq:emergencycontrol}) controls the system until a time when a satisfactory target set is found (at which point the MPC horizon resets to $N^{MPC}_k=T$).
The HPL algorithm in Sec.~\ref{sec:algorithm} ensures that whenever $N^{MPC}_k=0$, the system will be in the safe set (\ref{eq:emergencyset}) and the safety controller may be used. We prove this in Sec.~\ref{sec:feasproof}.

\section{The HPL Algorithm}\label{sec:algorithm}

Alg.~\ref{alg:HPL} summarizes the HPL control policy. 
Gaussian Processes, trained offline on trajectories from past control tasks, are used online to construct reduced-dimension strategy sets based on new environment forecasts. 
Target sets, computed by intersecting lifted strategy sets with the safety set, are used as terminal sets in a shifting-horizon MPC.

\begin{algorithm}
\caption{HPL Control Policy}\label{alg:HPL}
\begin{spacing}{1}
\begin{algorithmic}[1]
\State \textbf{\underline{parameters:}} $d_{\mathrm{thresh}}, T, N, \mathcal{X}_E, \pi_E$
\State \textbf{\underline{input:}} $f$, $\mathcal{X}$, $\mathcal{U}$, $\{\mathrm{Ex}(\mathcal{T}^1, {\bf{\Theta}}^1),..., \mathrm{Ex}(\mathcal{T}^n, {\bf{\Theta}}^n)\}, ~ {\bf{\Theta}}^{n+1}$
\State \textbf{\underline{output:}} $\mathrm{Ex}(\mathcal{T}^{n+1}, {\bf{\Theta}}^{n+1})$
\State
\State \textbf{\underline{offline:}} 
\State \textbf{train} GPs using stored executions as in Sec.\ref{ssec:trainingGP}
\State
\State \textbf{\underline{online:}} 
\State \textbf{initialize} $k=0$, $N^{MPC}_k = T$, $\mathrm{SetList} = [~]$
\For{each time step $k$}
\State \textbf{collect} $(x_k, \theta^{n+1}_{k:k+N})$
\State \textbf{find} $[\tilde{\mathcal{X}}_{k+T}, \tilde{\mathcal{U}}_{k : k+T}, \mathcal{C}_k]$ using (\ref{eq:gpeq}) - (\ref{eq:strategystateconstraint})
\If{$\mathcal{C}_k < d_{\mathrm{thresh}}$}
\State $\mathcal{X}_{k+T} = [~]$
\Else
\State  \textbf{construct} $\mathcal{X}_{k+T}$ using (\ref{eq:xt})
\EndIf
\State \textbf{append} $\mathcal{X}_{k+T}$ to $\mathrm{SetList}$ (\ref{eq:setlist}) and shift sets

\If{all sets in $\mathrm{SetList}$ (\ref{eq:setlist}) are \textbf{empty}}
\State $u_k = \pi_e(x_k, {\bf{\Theta}}^{n+1})$
\Else 
\State \textbf{calculate} $N_{k}^{MPC}$ using (\ref{eq:nk})
\State \textbf{solve} MPC (\ref{eq:lowlevelmpc}) with horizon $N_{k}^{MPC}$
\State $u_k = u_{0|k}^\star$ (\ref{eq:lowlevelinput})
\EndIf
\EndFor

\State \textbf{end}
\end{algorithmic}
\end{spacing}
\end{algorithm}

\section{Feasibility Proof}\label{sec:feasproof}
We prove that Alg.~\ref{alg:HPL} outputs a feasible execution for a new control task $\mathcal{T}^{n+1}$.

\textit{Theorem 1:}
Let Assumption 1 hold.
Consider the availability of feasible executions (\ref{eq:iterationvecs}) by a constrained system (\ref{eq:VehicleModel})-(\ref{eq:VehicleModelConstr}) of a series of control tasks $\{\mathcal{T}^1, \dots, \mathcal{T}^n \}$ in different environments $\{E({\bf{\Theta}}^1), \dots, E({\bf{\Theta}}^n)\}$.
Consider a new control task $\mathcal{T}^{n+1}$ in a new environment $E({\bf{\Theta}}^{n+1})$. 
If $x_0^{n+1} \in \mathcal{X}_E$, then the output of Alg.~\ref{alg:HPL} is a feasible execution of $\mathcal{T}^{n+1}$: $\mathrm{Ex}(\mathcal{T}^{n+1}, {\bf{\Theta}}^{n+1})$.

For ease of reading, we drop the task index $(\cdot)^{n+1}$.

\begin{proof}
We  use induction to prove that for all $k\geq 0$, the iteration loop (Lines 10-23) in Alg.~\ref{alg:HPL} finds an input $u_k$ such that the resulting closed-loop trajectory satisfies system and environment constraints. 

%First, we show that the HPL control policy is feasible at time step $k=0$ of the execution of the new task $\mathcal{T}^{n+1}$.
At time $k=0$ of the new task $\mathcal{T}^{n+1}$, the target set list can contain at most one non-empty set, $\mathcal{X}_T$ (\ref{eq:xt}). 
If $\mathcal{X}_T$ is non-empty, and the resulting (\ref{eq:lowlevelmpc}) is feasible, then there exists an input sequence $[u_{0|0}, \dots, u_{T-1|0}]$ calculated by (\ref{eq:lowlevelmpc}) satisfying all state and input constraints (\ref{eq:taskconstraints}), with $x_{T|0} \in \mathcal{X}_T$.
However, if $\mathcal{X}_T$ is empty or (\ref{eq:lowlevelmpc}) is infeasible, we instead apply the safety control law $u_0 = \pi_e(x_0, {\bf{\Theta}})$. By assumption, $x_0 \in \mathcal{X}_E$, so this input is feasible.
Thus we have shown that the iteration loop in Alg.~\ref{alg:HPL} is feasible for $k=0$.

Next, we show that the iteration loop of Alg.~\ref{alg:HPL} is recursively feasible.
Assume that at time $k>0$, the low-level policy (\ref{eq:lowlevelmpc}-\ref{eq:lowlevelinput}) is feasible with horizon $N^{MPC}_k$, and let ${\bf{x}}_{k:k+N^{MPC}_k | k}^{\star}$ and ${\bf{u}}_{k:k+N^{MPC}_k-1 | k}^{\star}$ be the optimal state trajectory and input sequence according to (\ref{eq:lowlevelmpc}), such that
\begin{align}
    u_k &= u_{k|k}^{\star}\label{eq:closedloopinput}\\
    x^{\star}_{k+N^{MPC}_k|k} &\in \mathcal{X}_{k+N^{MPC}_k}. \label{eq:stateintermconstraint}
\end{align}
If at time $k+1$ a non-empty target set $\mathcal{X}_{k+T+1}$ is constructed according to (\ref{eq:xt}) such that (\ref{eq:lowlevelmpc}) is feasible, then there exists a feasible input sequence $[u_{k+1|k+1}, \dots, u_{k+T|k+1}]$ satisfying all state and input constraints such that $x_{k+T+1} \in \mathcal{X}_{k+T+1}$. 

If at time $k+1$ the target set is \textit{empty}, or (\ref{eq:lowlevelmpc}) is infeasible, we must consider two cases separately:

\underline{Case 1: The MPC horizon at time step $k$ is $N^{MPC}_k > 1$.}
In the absence of model uncertainty, when the closed-loop input $u_k$ (\ref{eq:closedloopinput}) is applied, the system (\ref{eq:VehicleModel}) evolves such that
\begin{align}\label{eq:statevolution}
    x_{k+1} = x^{\star}_{k+1|k}.
\end{align}
According to Alg.~\ref{alg:HPL}, when the empty target set $\mathcal{X}_{k+1+T}$ is added to the target set list (\ref{eq:setlist}), the MPC horizon is shortened and the most recent non-empty target set is used again. Since $N^{MPC}_k > 1$, we are guaranteed at least one non-empty target set in (\ref{eq:setlist}) that may used as a feasible terminal constraint in the low-level controller (\ref{eq:lowlevelmpc}).
At time step $k+1$, the shifted input sequence ${\bf{u}}_{k+1:k+N_k^{MPC}-1 | k}^{\star}$ will be optimal for this shifted horizon optimal control problem (with a corresponding state trajectory ${\bf{x}}_{k+1:k+N_k^{MPC} | k}^{\star}$). At time step $k+1$, Alg.~\ref{alg:HPL} applies the second input calculated at the previous time step: $u_{k+1} = u^{\star}_{k+1 | k}$.

\underline{Case 2: $N^{MPC}_k = 1$.} In this scenario, the target set list (\ref{eq:setlist}) at time step $k+1$ is empty, resulting in $N^{MPC}_{k+1} = 0$ (\ref{eq:nk}).
However, combining the fact that $N^{MPC}_k=1$ with (\ref{eq:stateintermconstraint}) and (\ref{eq:statevolution}), we note that
\begin{align}
    x_{k+1} \in \mathcal{X}_{k+1} \subseteq \mathcal{X}_E, \nonumber
\end{align}
by construction of the target set (\ref{eq:xt}). 
This implies that at time step $k+1$, system (\ref{eq:VehicleModel}) is necessarily in the safe set (\ref{eq:emergencyset}), and so application of the safety controller (\ref{eq:emergencycontrol}) will result in a feasible input, $u_{k+1} = \pi_e(x_{k+1},{\bf{\Theta}})$.

We have shown that \textit{i)} the online iteration loop (Lines 10-23) in Alg.~\ref{alg:HPL} finds a feasible input at time step $k=0$, and \textit{ii)} if the loop finds a feasible input at time step $k$, it must also find a feasible input at time $k+1$. 
We conclude by induction that Alg.~\ref{alg:HPL} finds a feasible input $u_k~ \forall k \in \mathbb{Z}_{0+}$ in the new task $\mathcal{T}^{n+1}$. This results in a feasible execution of $\mathcal{T}^{n+1}$.
\end{proof}

\section{Robotic Manipulator Navigation}\label{sec:robot}
We evaluate HPL in a robotic path planning example. 

\begin{figure}
    \centering
    \includegraphics[width=\columnwidth]{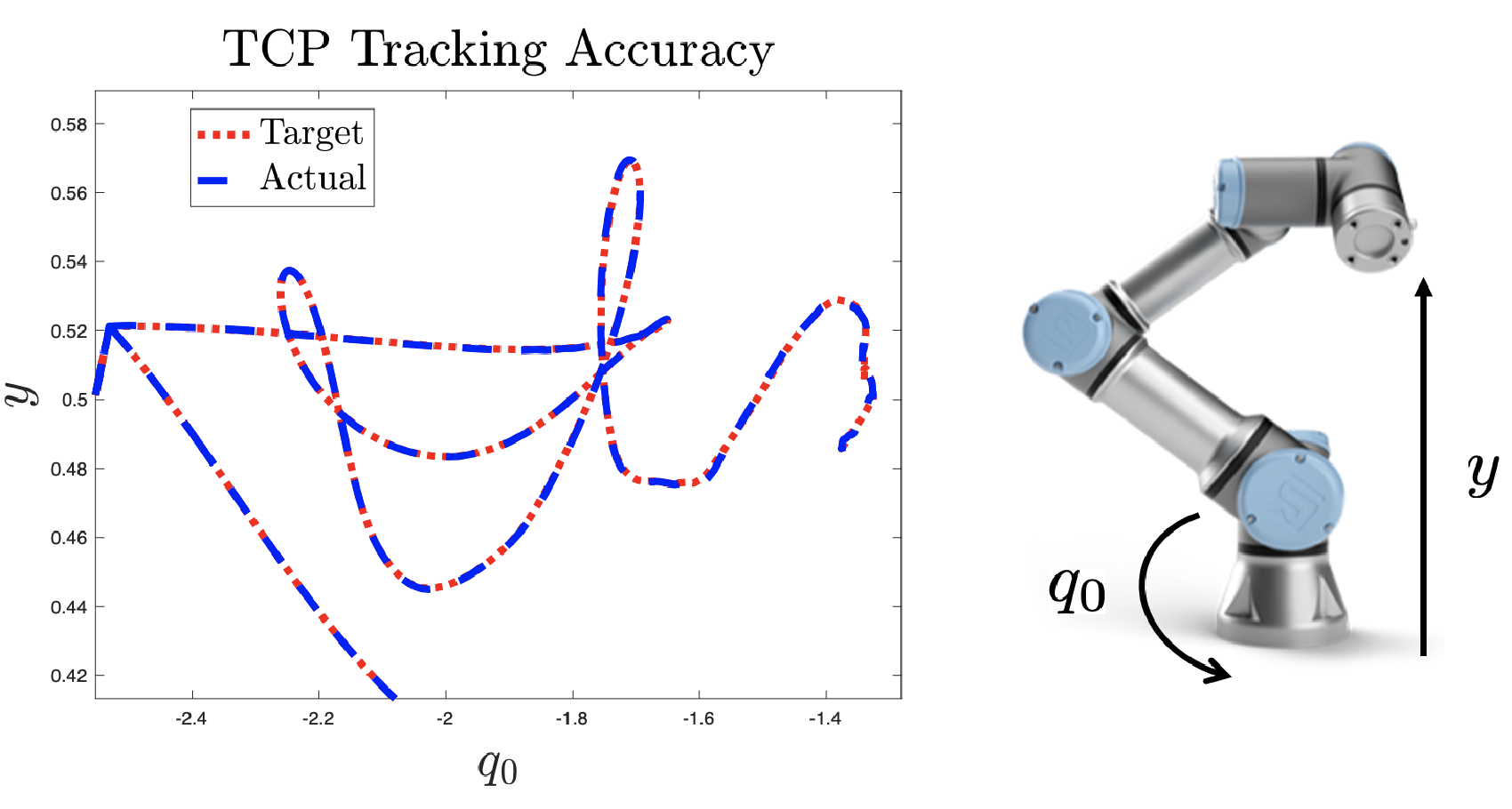}
    \caption{The high TCP tracking accuracy of the UR5e robot manipulator allows for use of a simplified end-effector model in $\{q_0, y\}$ space.}
    \label{fig:ur5tracking}
\end{figure}

\subsection{System and Task description}

Consider a UR5e\footnote{https://www.universal-robots.com/products/ur5-robot/} robotic arm tasked with maneuvering its end-effector along a tube with varying slope.
The UR5e has high end-effector reference tracking accuracy (as depicted in Fig.~\ref{fig:ur5tracking}), allowing us to use a simplified end-effector model in place of a discretized second-order model \cite{robothard3}.
At each time step $k$, the state of the system is $x_k$,
\begin{align}
    x_k = [q_{0_k},~ \dot{q}_{0_k},~y_k, ~\dot{y}_k], \nonumber
\end{align}
where $q_{0_k}$ is the robot's first horizontal joint angle, $y_k$ the height of the end-effector, and $\dot{q}_{0_k}$ and $\dot{y}_k$ their respective velocities (see Fig.~\ref{fig:ur5tracking}).
The inputs to the system are 
\begin{align}\label{eq:systeminputs}
    u_k = [\ddot{q}_{0_k},~ \ddot{y}_k],
\end{align}
the accelerations of the end-effector in the $q_0$ and $y$ direction, respectively.
We model the base-and-end-effector system as a quadruple integrator:
\begin{align}
    {x}_{k+1} & = A x_k + B u_k \label{eq:simplifiedRobotModel}\\ 
    A &=
    \begin{bmatrix}1 & dt & 0 & 0 \\ 0 & 1 & 0 & 0\\ 0 & 0 & 1 & dt \\ 0 & 0 & 0 & 1 \end{bmatrix}, B = \begin{bmatrix}0 & 0 \\ dt & 0 \\ 0 & 0 \\ 0 & dt \end{bmatrix}, \nonumber
\end{align}
where $dt = 0.01$ seconds is the sampling time. 
This model accurately represents the system dynamics as long as we operate within the region of high end-effector reference tracking accuracy, characterized experimentally as the following state and input constraints:
\begin{equation}\label{eq:taskinputconstr}
\begin{aligned}
    {\mathcal{X}} &=  \begin{bmatrix}-3 \\ -3\end{bmatrix} \leq  \begin{bmatrix}\dot{q}_{0_k} \\ \dot{y}_k  \end{bmatrix}  \leq   \begin{bmatrix}3  \\ 3 \\ \end{bmatrix}  \nonumber \\
    {\mathcal{U}} &=  \sqrt{\ddot{q}_{0_k} + \ddot{y}_k} \leq 1, \nonumber
\end{aligned}
\end{equation}
where the states $q_{0_k}$ and $y_k$ are not constrained by the system, but by the particular task environment.

Each control task $\mathcal{T}^i$ requires the end-effector to be controlled through a different tube, described using the environment descriptor function ${\bf{\Theta}}^i$, as quickly as possible.
Here, the function ${\bf{\Theta}}^i$ maps a state in a tube segment to the slope of the constant-width tube. Different control tasks $\{\mathcal{T}^1, \dots, \mathcal{T}^n \}$ correspond to maneuvering through tubes of constant width but different piecewise-constant slopes.

We choose two strategy states for these tasks: 
\begin{align}
    \tilde{x}_k = [s_k, ~h_k], \nonumber
\end{align}
where $s_k$ is the cumulative distance along the centerline of the tube from the current point $(q_{0_k}, y_k)$ to the projection onto the centerline, and $h_k$ is the distance from $(q_{0_k}, y_k)$ to the centerline. The strategy states therefore measure the total distance traveled along the tube up to time step $k$, and the current signed distance from the center of the tube. 
These strategy states were chosen because they provide information about both task performance (distance traveled along the tube is a measure of task completion speed) and constraint satisfaction (distance from the tube boundaries). Note that in this task, just as in car racing, cutting corners (as measured by $h_k$) also maximizes distance traveled along the tube. 
% The change-of-coordinate function $g$ for this system is given by
% \begin{align}
%     g(x_k, y_k) & =  [s_k,~ h_k] \\
%     &= [\mathrm{proj}_C(x_k, y_k), \pm ||(x_k, y_k) - \mathrm{proj}_C(x_k, y_k)||], \nonumber
% \end{align}
% where $\mathrm{proj}_C(\cdot, \cdot)$ is the projection onto the tube centerline. The sign of $h_k$ is positive if the end-effector is above the tube centerline.
% The strategy state space is given by
% \begin{equation}
%   \begin{aligned}
%     \tilde{\mathcal{X}} &=  \begin{bmatrix}0 \\ -\frac{w}{2}\end{bmatrix} \leq  \begin{bmatrix}s_k \\ h_k  \end{bmatrix}  \leq   \begin{bmatrix}l  \\ \frac{w}{2} \\ \end{bmatrix},
% \end{aligned} 
% \end{equation}
% where $l$ is the total length of the tube and $w$ the width. The task setup and strategy state coordinates are depicted in Fig.~\ref{fig:coordinates}.
The system inputs (\ref{eq:systeminputs}) are used as strategy inputs.
\begin{figure}
    \centering
    \includegraphics[width=\columnwidth]{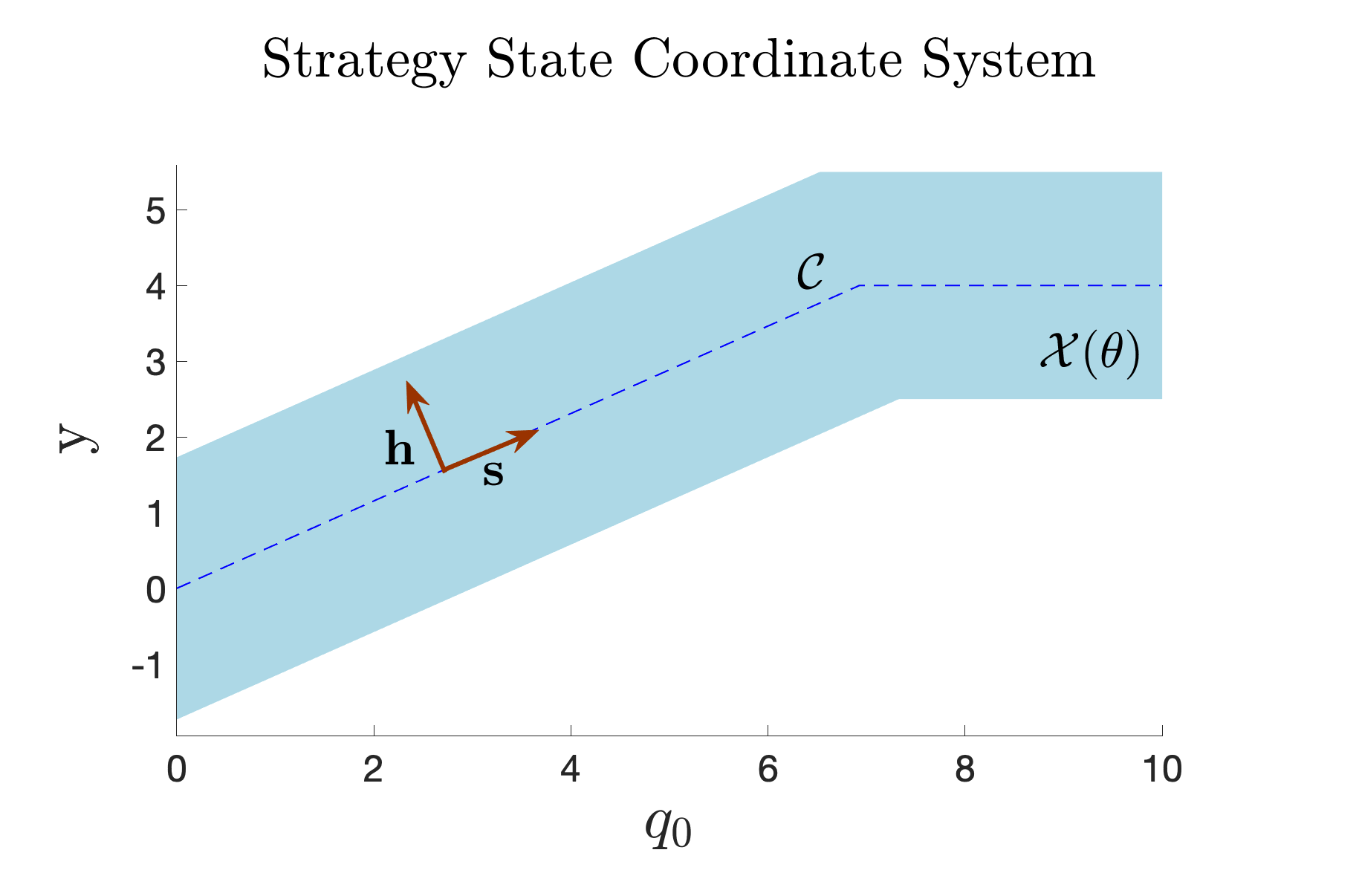}
    \caption{The end-effector is constrained to stay in the light blue tube $\mathcal{X}(\bf{\Theta})$. The strategy states measure the cumulative distance along and the distance from the centerline.}
    \label{fig:coordinates}
\end{figure}

The safety controller (\ref{eq:emergencycontrol}) is an MPC controller which tracks the centerline of the tube at a slow, constant velocity of $0.5$ meters per second. 
%The system (\ref{eq:simplifiedRobotModel}) in closed-loop with this centerline-tracking controller is able to solve each of the considered tasks without breaking the state constraints imposed by the environment (i.e. without hitting the tube boundary).
The safe set $\mathcal{X}_E$ (\ref{eq:emergencyset}) is determined offline using sampling-based forward reachability.

\subsection{Hierarchical Predictive Learning Results}
\begin{figure}
    \centering
    \includegraphics[width=\columnwidth]{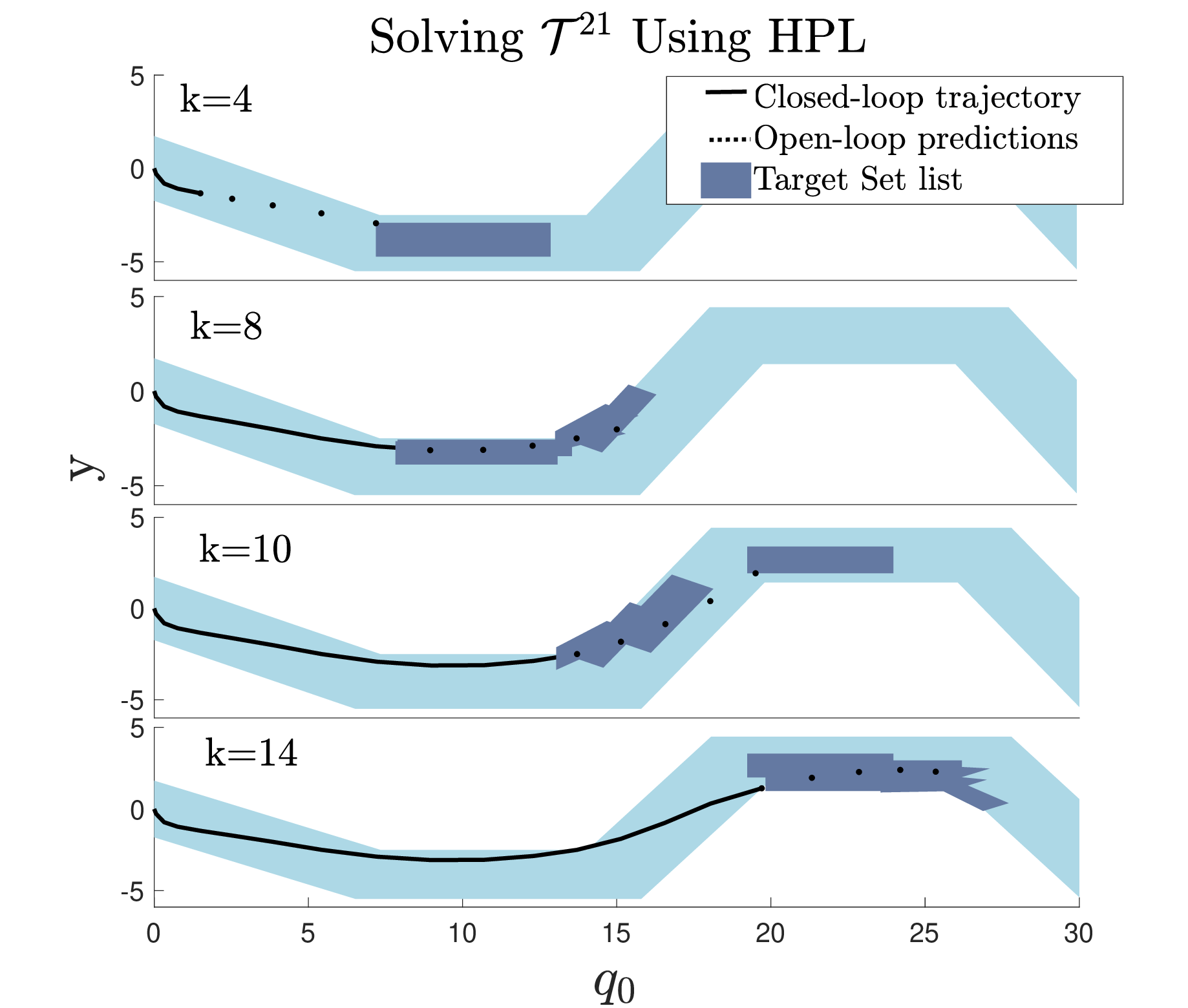}
    \caption{At each time step, the target set list (\ref{eq:setlist}) provides different regions in the task space for the system to track. %\textit{Note:} The apparent constraint violation by the raceline is due to the system discretization.
    }
    \label{fig:plotprogression}
\end{figure}

\begin{figure}
    \centering
    \includegraphics[width=\columnwidth]{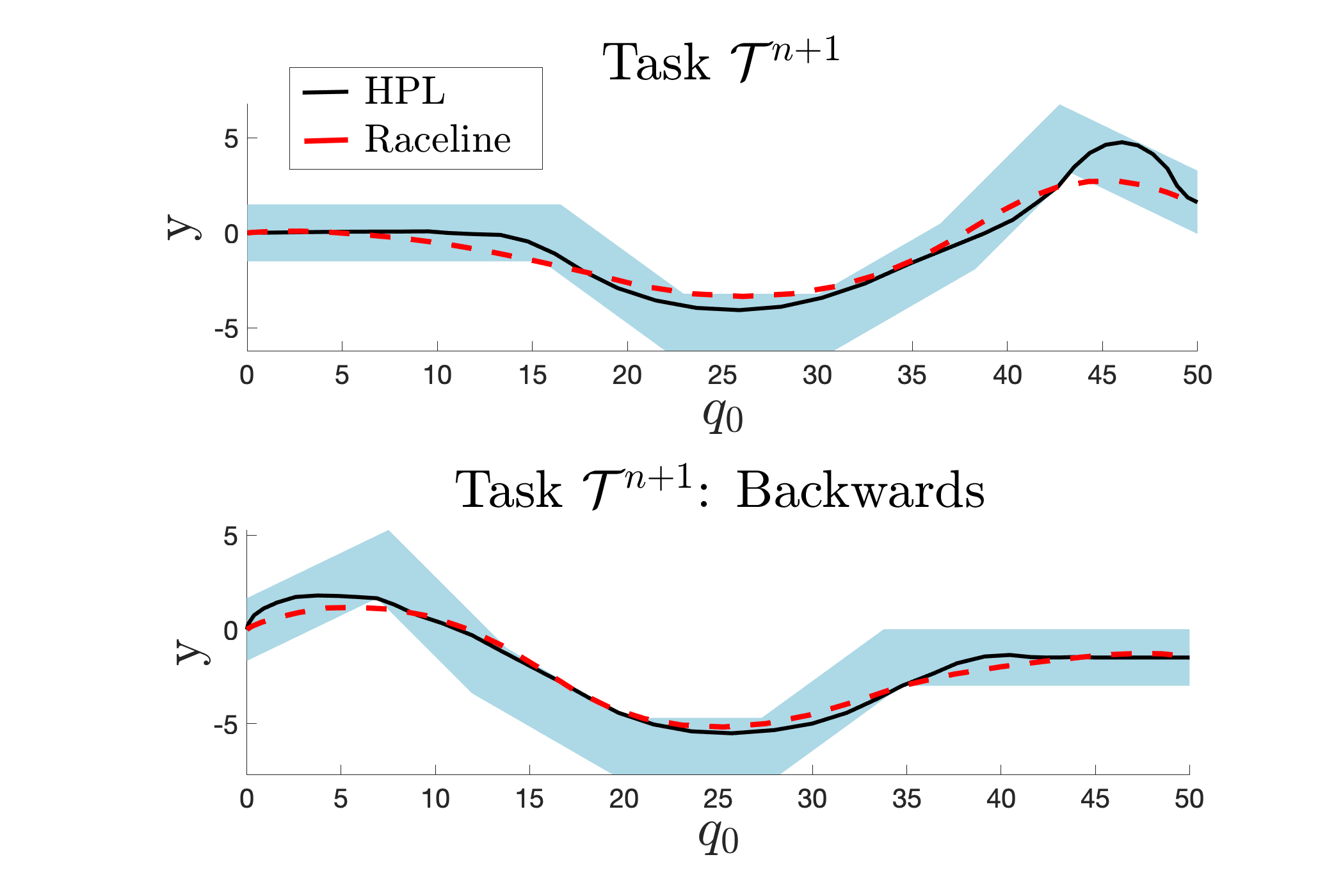}
    \caption{The HPL execution is compared to the raceline (the fastest possible execution), as determined by an LMPC \cite{rosolia2016learning}. Respective execution times in [s] are $6.5$ (LMPC), $8.8$ (HPL), and $12.8$ (Centerline-tracking $\pi_e$).}
    \label{fig:bf1}
\end{figure}
We test the effectiveness of the HPL control architecture (Alg.~\ref{alg:HPL}) in simulation.
We begin by collecting executions that solve a series of $20$ control tasks $\{ \mathcal{T}^1, ...,\mathcal{T}^{20}\}$, with each control task corresponding to a different tube shape. 
The executions are closed-loop trajectories completed by a Learning Model Predictive Controller (LMPC). This reference-free iterative learning controller is initialized with a conservative, feasible trajectory and then improves its closed-loop performance at each iteration of a task. For each control task $\mathcal{T}^i$, we find an initial (suboptimal) execution using the centerline-tracking MPC safety controller. This execution is used to initialize the LMPC, which then runs for five iterations on the task $\mathcal{T}^i$. At each iteration, the LMPC uses the trajectory data from the previous iteration to improve the closed-loop performance at the current iteration with respect to a chosen cost function (in our instance, time required to reach the end of the tube). 
The execution corresponding to the fifth LMPC iteration of each task $\mathcal{T}^i$ is added to our training data set.

We use an environment forecast horizon of $N=10$ seconds and a control horizon of $T=5$ seconds.
These horizons were chosen based on the approximate required time to traverse an average tube segment in the previous tasks. The forecast horizon was chosen to be twice the control horizon in order to provide information about both the current and subsequent tube segments.
GPs using the squared-exponential kernel are then optimized to approximate the strategies learned from solving the $20$ different control tasks. 
Matlab 2018a was used for all data collection and GP training.

Figure \ref{fig:plotprogression} shows the closed-loop trajectory and target set list at various time steps of solving a new task, $\mathcal{T}^{21}$, using the HPL framework. At each time step, the final predicted state lies within the last set in the target set list; the other predicted states track any non-empty target sets as closely as possible. 
%As is clear from Fig.~\ref{fig:plotprogression}, some FDS sets in the FDS list are empty at certain timesteps. 
The formulation allows us to visualize what strategies have been learned, by plotting at each time step where the system thinks it should go. Indeed, we see that the system has learned to maneuver along the insides of curves, and even takes the direct route between two curves going in opposite directions.

Figure \ref{fig:bf1} shows the resulting executions for a new task solved forwards and backwards. 
We emphasize that HPL generalizes the strategies learned from training data to unseen tube segments. 
Specifically, for the tasks shown here the GPs were trained on executions solving tasks in the forward direction, i.e. constructed left to right using tube segments as shown in the top images.
%When the tasks are solved backwards, the environment descriptors are piecewise mirror images of previously seen environment descriptors. 
For example, tube segments of certain slopes had only been traversed upwards in previous control tasks, never down.
The HPL control architecture was able to handle this change very well. 
As in Fig.~\ref{fig:plotprogression}, the forward and backwards trajectories demonstrate good maneuvering strategies, including moving along the insides of curves and cutting consecutive corners.
In fact, the HPL controller finds an execution that is very close to the minimum time execution as determined by an iterative learning MPC controller \cite{ugoproof}.

\subsection{Implementation details}
\begin{figure}
    \centering
    \includegraphics[width=0.9\columnwidth]{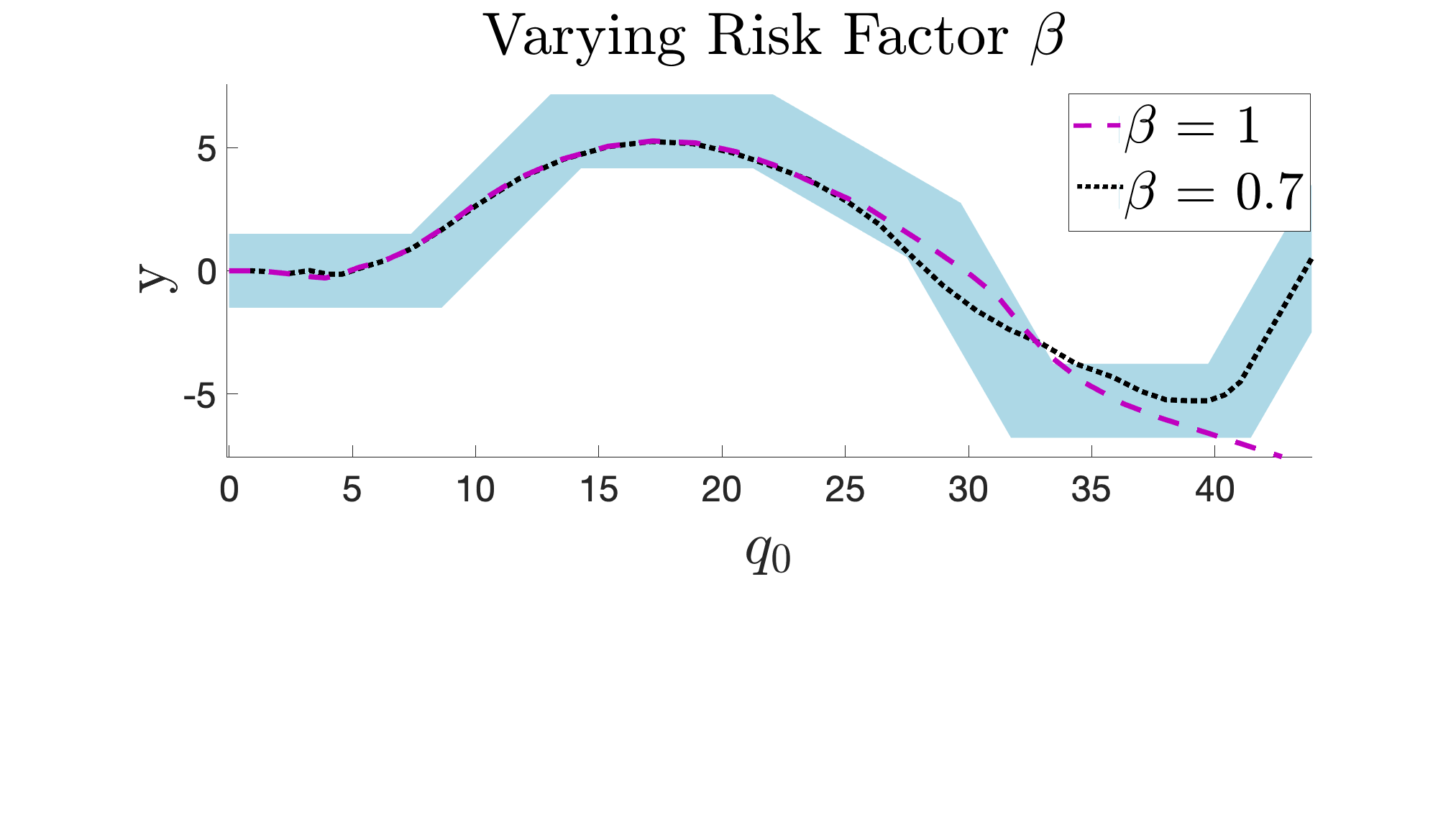}
    \caption{The same task is performed using two different values of $\beta$. For this task, using $\beta = 1$ led to eventual constraint violation because the terminal constraints were not formed in conjunction with the safety controller.}
    \label{fig:traj_beta_comp}
\end{figure}

Section~\ref{ssec:betarisk} introduced an approach for tuning the conservativeness with which the strategy sets (\ref{eq:strategystateconstraint}) are lifted into target sets (\ref{eq:xt}), by varying the importance of intersecting with the safety set $\mathcal{X}_E$ (\ref{eq:emergencyset}).
To examine the effects of varying the acceptable risk level $\beta$ on the system's closed-loop trajectory, we solve a single task using $\beta = 1$ and a more conservative $\beta = 0.7$.
The results are depicted in Fig.~\ref{fig:traj_beta_comp}, and demonstrate the benefits of incorporating the safety set $\mathcal{X}_E$ during the formation of full-dimensional target sets.

Both trajectories ($\beta = 0.7$ and $\beta = 1$) follow the same path at the beginning of the task. However, during a sharp curve towards the latter portion of the task, the trajectory corresponding to $\beta = 1$, plotted in purple, is no longer able to satisfy the task constraints. 
%When $\beta = 1$, the constraints imposed by the safety controller are not taken into account when forming the target set, and therefore the system moves into parts of the state space where the safety controller cannot recover the system. 

The more conservative approach, using $\beta = 0.7$ and plotted in black, takes the safety constraints $\mathcal{X}_E$ into account and is able to maneuver the sharp turn. We see that the center-tracking safety controller activates at the end of the task and keeps the system within the constraints (whereas the $\beta = 1$ trajectory leaves the task constraints at this point). 

\section{Formula 1 Racing}\label{sec:form1}
We evaluate HPL control in an autonomous racing task, where a small RC car is controlled as quickly as possible around $1/10$ scale Formula 1 race tracks. 

\subsection{System and Task Description}\label{sec:racing_taskdesc}
Consider a small RC car\footnote{\href{http://www.barc-project.com/}{\texttt{http://www.barc-project.com/}}} whose dynamics are modeled in the curvilinear abscissa reference frame \cite{rajamani2011vehicle} with a nonlinear Pacejka tire model, with states and inputs at time step $k$:
\begin{align}
    x_k & = [v_{x_k} ~ v_{y_k} ~ \dot\psi_{k} ~ e_{\psi_k} ~ s_k ~ e_{y_k} ]^\top, \label{eq:barcmodel}\\
    u_k &= [a_{k} ~ \delta_k ]^\top, \nonumber
\end{align}
where $v_{x_k}$, $v_{y_k}$, and $\dot{\psi}_{k}$ are the vehicle's longitudinal velocity, lateral velocity, and yaw rate, respectively, at time step $k$, $s_k$ is the distance traveled along the centerline of the road, and $e_{\psi_k}$ and $e_{y_k}$ are the heading angle and lateral distance error between the vehicle and the path. The inputs are longitudinal acceleration $a_k$ and steering angle $\delta_k$. 

The system dynamics are described using an Euler discretized dynamic bicycle model \cite{rosolia2017autonomousrace}. The vehicle is subject to system-imposed state and input constraints given by
\begin{equation}
    \mathcal{X} = \left\{ x : \begin{bmatrix}0 \\ -10~\mathrm{m/s} \\ -\frac{\pi}{2}~ \mathrm{rad}\\ -\frac{\pi}{3} ~\mathrm{rad} \\ -\frac{l}{2}~ \mathrm{m}\end{bmatrix} \leq  \begin{bmatrix}v_x \\ v_y\\ w_z \\ e_{\psi} \\ e_y \end{bmatrix}  \leq   \begin{bmatrix}10~\mathrm{m/s} \\ 10~\mathrm{m/s}\\ \frac{\pi}{2} ~\mathrm{rad}\\\frac{\pi}{3} ~\mathrm{rad}\\  \frac{l}{2}~ \mathrm{m} \end{bmatrix} \right \}, \label{eq:car_X}
\end{equation}
\begin{equation}
    \mathcal{U} = \left\{ u : \begin{bmatrix}-1~{\mathrm{m/s^2}} \\ -0.5~{\mathrm{rad/s^2}} \end{bmatrix} \leq  \begin{bmatrix}a \\ \delta \end{bmatrix}  \leq   \begin{bmatrix}1~{\mathrm{m/s^2}} \\ 0.5~{\mathrm{rad/s^2}} \end{bmatrix} \right \}, \nonumber
\end{equation}
where $l = 0.8 $ is the track's lane width.

The car's task is to drive around a race track as quickly as possible while satisfying all system and environmental state and input constraints. Each control task $\mathcal{T}^i$ corresponds to a new track, described using the environment descriptor function ${\bf{\Theta}}^i$ which maps the current position along the $i$-th track to a description of the upcoming track curvature. 
For training and testing, we use scaled Formula 1 tracks whose geospatial coordinates were made publicly available in \cite{belien}.

We choose two similar strategy states for the racing task as for  the robot navigation task: 
\begin{align}
    \tilde{x}_k = [\Delta s_k, ~e_{y_k}], \label{eq:barcstratstate}
\end{align}
where $\Delta s_k = s_k - s_{k-T}$ measures the distance traveled along the track centerline in the last $T$ timesteps. 
%The future strategy states $\tilde{x}_{k+T}$ predicted by the learned GP at time $k$ thus measure how far the vehicle should travel along the track in the next $T$ time steps, and the final deviation from the centerline at time $k+T$.

The safety controller is an MPC controller which tracks the centerline of the race track at a constant velocity of $5~ \mathrm{m/s}$. 
The safe set $\mathcal{X}_E$ induced by this safety controller is estimated from Monte Carlo simulations. 
%A set of initial states are uniformly sampled from $\mathcal{X}$ (\ref{eq:car_X}), and labeled according to the feasibility (or infeasibility) of each resulting $2N_{MPC}$-step closed-loop trajectory using the safety MPC controller. A polytopic approximation of the true safe set $\mathcal{X}_E$ is determined by fitting an SVM classifier \cite{cortes1995support} with linear kernel to the labeled initial conditions. 

\subsection{Hierarchical Predictive Learning Results}
We construct ten different $1/10$-scale Formula 1 tracks from \cite{belien}, and then calculate the minimum-time trajectory for each track using a continuous-time vehicle model. 
The discretized racelines from seven tracks are used to create training data for our strategy, while the remaining three tracks are used for evaluating the performance of the HPL control algorithm. 

GPs using the squared-exponential kernel are trained on these racelines for each of the two strategy states. As described in Sec.~\ref{ssec:stratHor}, the environment forecast does not have to be parameterized by time. Indeed, for the racing task we use an environment forecast parameterized by a constant distance along the track centerline, regardless of the vehicle's velocity:
\begin{align*}
    \theta^i_{k:k+N} = [{\bf{\Theta}}^i(s^i_{k}),{\bf{\Theta}}^i(s^i_{k} + 2) \dots, {\bf{\Theta}}^i(s^i_{k} + 2N)],
\end{align*}
where $s^i_k$ is the total distance traveled along the centerline of the track at time $k$. 
Given the vehicle's current state (\ref{eq:barcmodel}) and this environment forecast, the learned GP strategy predicts the target values of the strategy states (\ref{eq:barcstratstate}) at $T=2$ seconds into the future. 
The forecast and control horizon were chosen by examining past task data and estimating how close an environment change had to be in order to impact the vehicle's locally optimal trajectory.
Training two GPs using GPyTorch on a 2017 MacBook Pro with 2.8 GHz Quad-Core Intel Core i7 took $494$ seconds. 
Evaluating these GPs took an average of $0.004$ seconds on the same processor. 

\begin{figure}
    \centering
    \includegraphics[width=0.85\columnwidth]{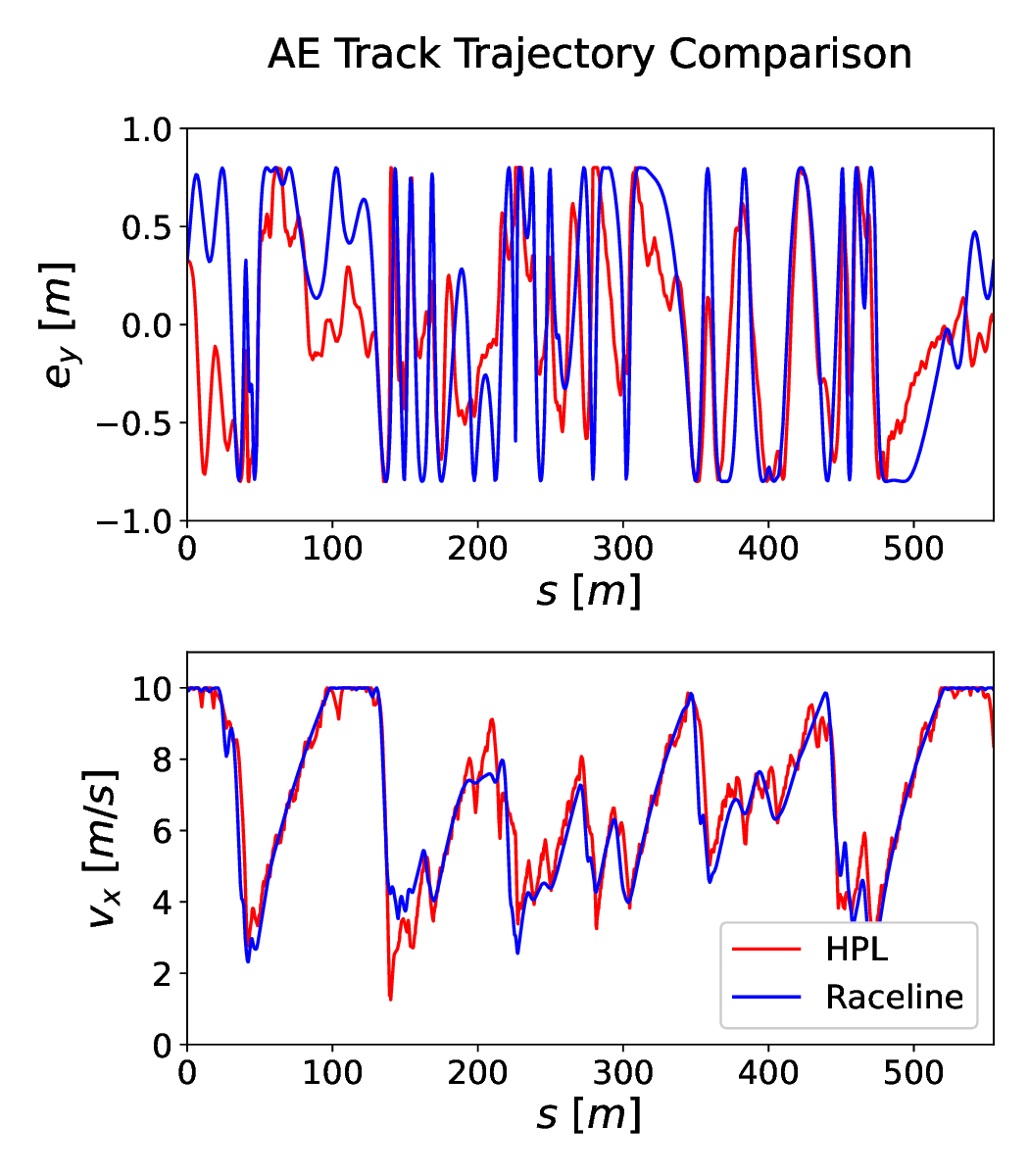}
    \caption{The two strategy GPs, trained using minimum-time trajectory data from seven race tracks, are able to predict (red) the centerline deviation and longitudinal velocity of a new, unseen track's raceline (blue).}
    \label{fig:AE_trajcomp}
\end{figure}

Figure~\ref{fig:AE_trajcomp} compares two states ($v_x$ and $e_y$) of the HPL closed-loop trajectory on a new task (the ``AE" track from \cite{belien}) with those of the optimal minimum-time trajectory. These two states are most informative on whether a good strategy was learned, since they correspond most closely with the chosen strategy states. The learned strategy performs well on this new task, and is able to pick up the pattern between track curvature and the raceline.
the GP trained to predict how far the vehicle should travel in the next $T$ steps (affecting the resulting $v_x$ values) performs slightly better than the GP trained to predict deviation from the centerline ($e_y$). This trend was found in all three test tracks, and is likely due to deviations in optimal steering on straightaways, where the angle of the raceline depends on a longer environment forecast than considered.

\begin{figure}
    \centering
    \includegraphics[width=0.95\columnwidth, trim = {6.1cm 0 1.2cm 0.2cm}, clip]{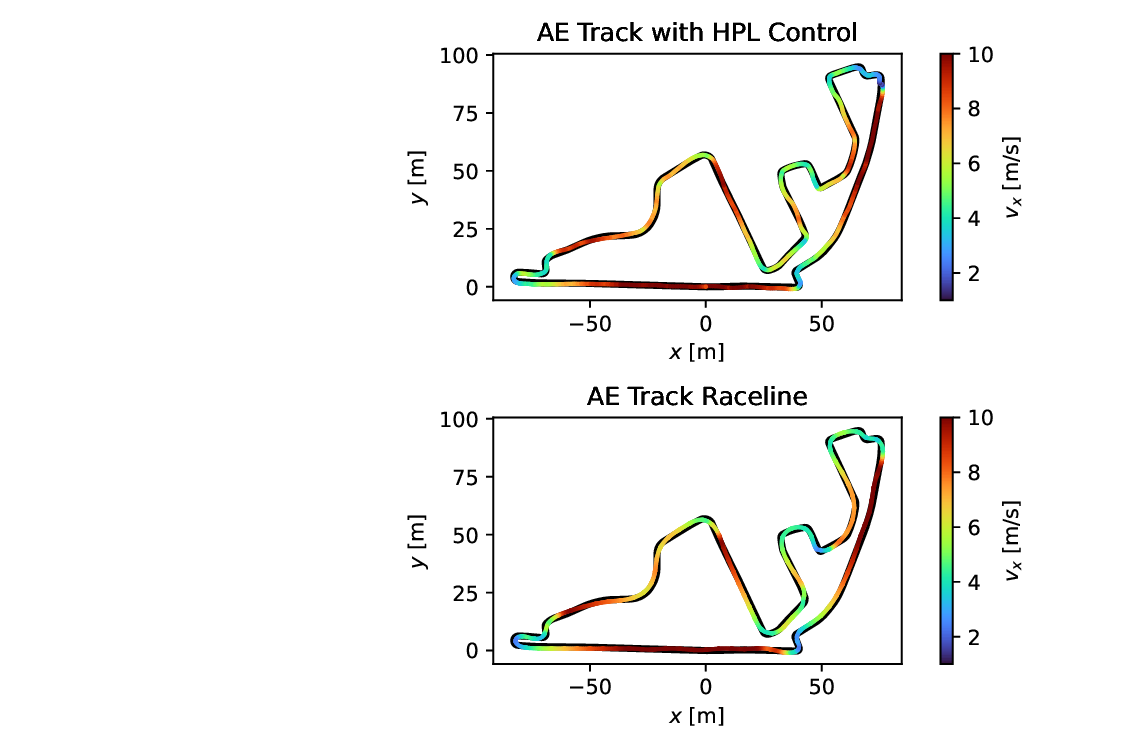}
    \caption{The HPL controller is able to match the speed profile and shapes of the minimum time trajectory, slowing down in curves and speeding up to the maximum allowable velocity on straight segments.}
    \label{fig:AE_velcomp}
\end{figure}

The HPL trajectory is plotted and compared to the track's raceline in Fig.~\ref{fig:AE_velcomp}, with a colormap indicating how quickly the car drives in various sections of the track. Confirming the trend in Fig.~\ref{fig:AE_trajcomp}, the plot shows the HPL control accurately predicts the ideal vehicle speed, speeding up and slowing down in the same sections of the racetrack as the optimal raceline. It is also clear that the system has learned various smart driving rules introduced in Sec.~\ref{ssec:racerules}, including steering out before cutting the insides of corners and taking direct routes between two curves.
More examples of the HPL closed-loop trajectory, with snippets from all three test tracks, are depicted in Fig.~\ref{fig:trackExamples}, and further plots are included in the Appendix.

The HPL control architecture results in feasible and fast executions for all three tested race tracks. Because the learned strategy GPs were able to predict the velocity trend so well (see Fig.~\ref{fig:AE_trajcomp}), the three task durations were all within 5\% of the respective raceline time (see Table~\ref{tab:racetimetable}). We again stress the ability of the strategy GPs to extrapolate patterns seen during training despite the tracks' very varied curvature.

In order to evaluate the benefit of using a hierarchical control framework, we also tested a controller with higher risk factor. 
As described in Sec.~\ref{ssec:betarisk}, this type of controller ($\beta = 1$) does not take the safety set $\mathcal{X}_E$ into account when creating the target set. As a result, the MPC terminal set constrains the strategy states to be in the GP's predicted set and the remaining state to simply be in the allowable state space. The safety set $\mathcal{X}_E$ is not used to guide the process of lifting the reduced-dimension strategy set into the full-dimensional target set. This means there are no guarantees that predicted terminal state will be in the domain of our safety control, and therefore it is possible to lose feasibility later in the task. 
Indeed, in the racing task (just as in the robotic manipulation task in Sec.~\ref{sec:robot}), the simple high-risk controller was unable to complete a single lap without becoming infeasible. These failures occurred early in each of the three test tracks (in the first $10\%$ of the tracks), during sharp corners where the safety controller was required but infeasible. 
This again demonstrates the value of using both data-driven and physics-based components, and a hierarchical framework for integrating them in a structured manner.

\hspace{1pt}
\begin{center}
\begin{tabular}{ |l | c  c  | } 
\hline
  Track & HPL Lap Time [s] & Raceline Lap Time [s] \\
   \hline
 AE  & 87.6 & 85.1 \\ 
BE & 92.3 & 89.3 \\ 
 US & 85.6 & 83.3 \\ 
 \hline
\end{tabular}
\captionof{table}{The HPL controller results in lap times less than 5\% longer than the minimum-time trajectory, demonstrating that an effective racing strategy was learned.}\label{tab:racetimetable}
\end{center}
\hspace{1pt}

\begin{figure}[htp]
\centering 
\subfloat[]{%
  \includegraphics[width=0.9\columnwidth]{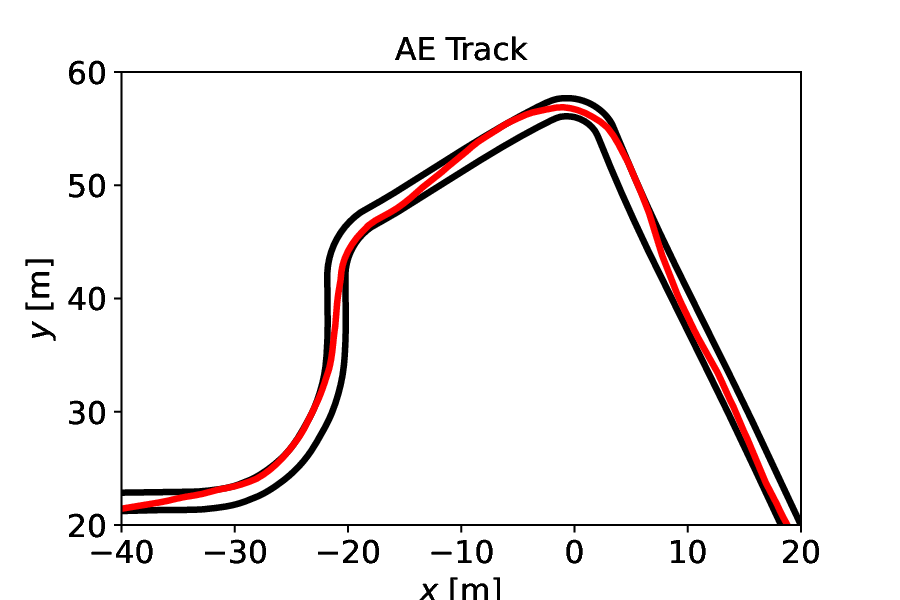}%
}

\subfloat[]{%
  \includegraphics[width=0.9\columnwidth]{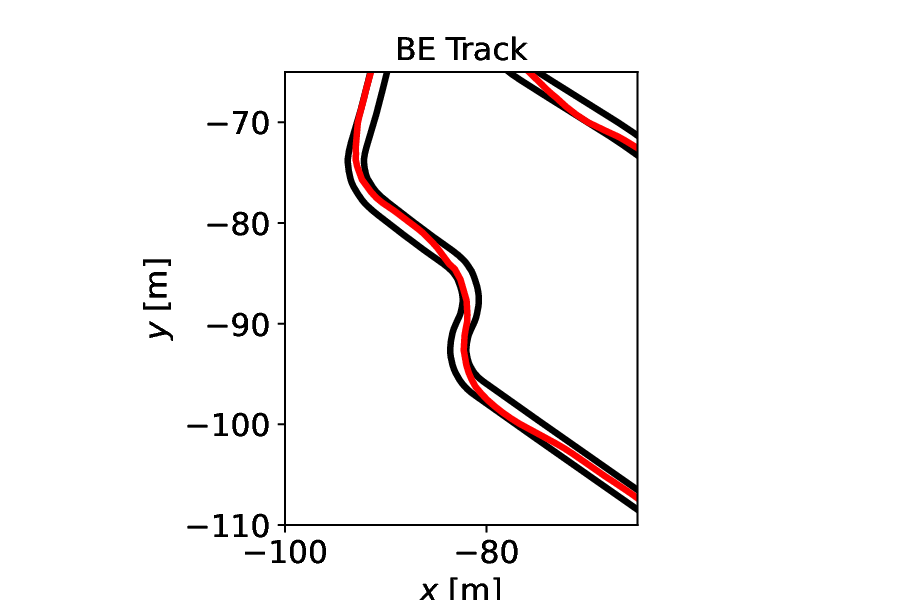}%
}

\subfloat[]{%
  \includegraphics[width=0.9\columnwidth]{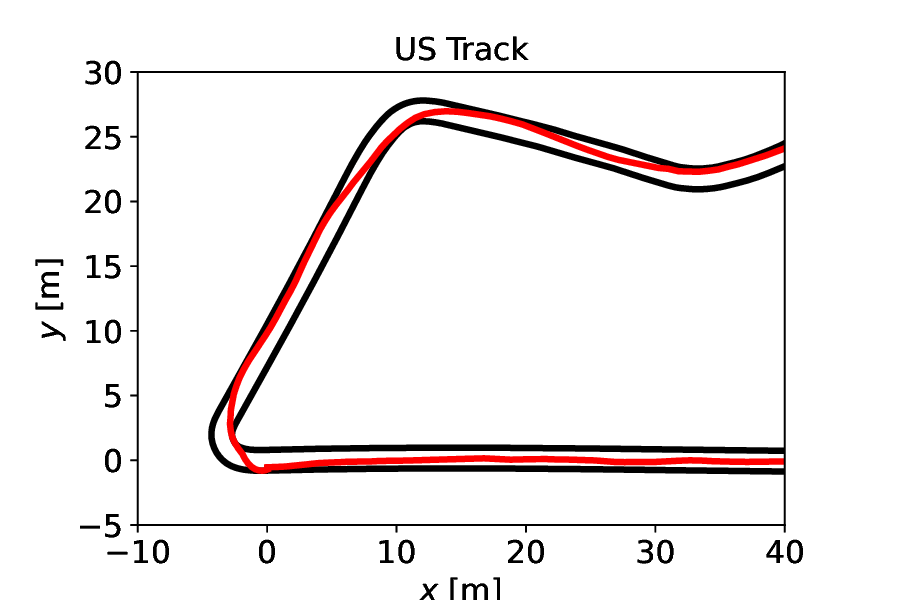}%
}

\caption{The HPL controller uses learned strategies to cut corners in all three test race tracks.}
\label{fig:trackExamples}

\end{figure}

\subsection{Neural Network as Strategy}
An additional trial was run where a neural network, rather than a GP, was used to represent the strategy.
The training data was constructed in the same way as described in Sec.~\ref{sec:racing_taskdesc}, and a neural network with five linear hidden layers was trained using the python SKLearn library. One network could be trained to predict both strategy states, requiring $15$ seconds of training time using Torch on a Pro 2017 MacBook Pro with 2.8 GHz Quad-Core Intel Core i7. 

\begin{figure}
    \centering
    \includegraphics[width=\columnwidth, trim = {5cm 0 1.2cm 0.2cm}, clip]{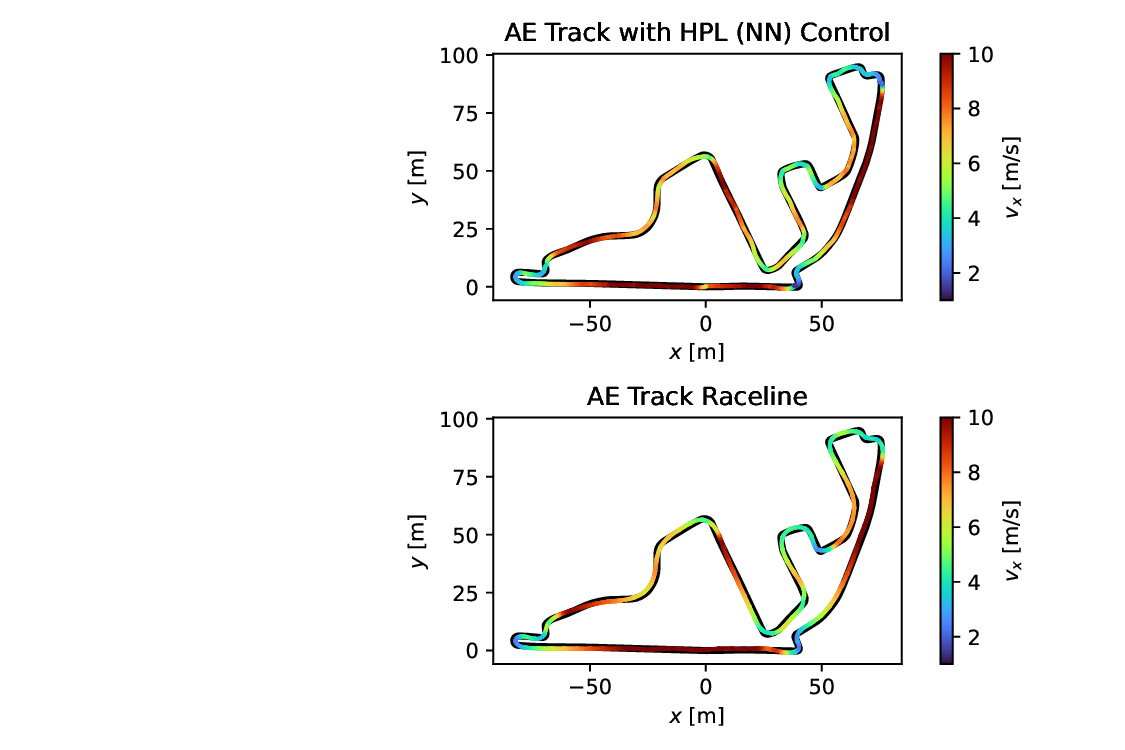}
    \caption{The HPL controller is able to match the speed profile and shapes of the minimum time trajectory }
    \label{fig:nn_AE_vel}
\end{figure}

\begin{figure}
    \centering
    \includegraphics[width=0.8\columnwidth]{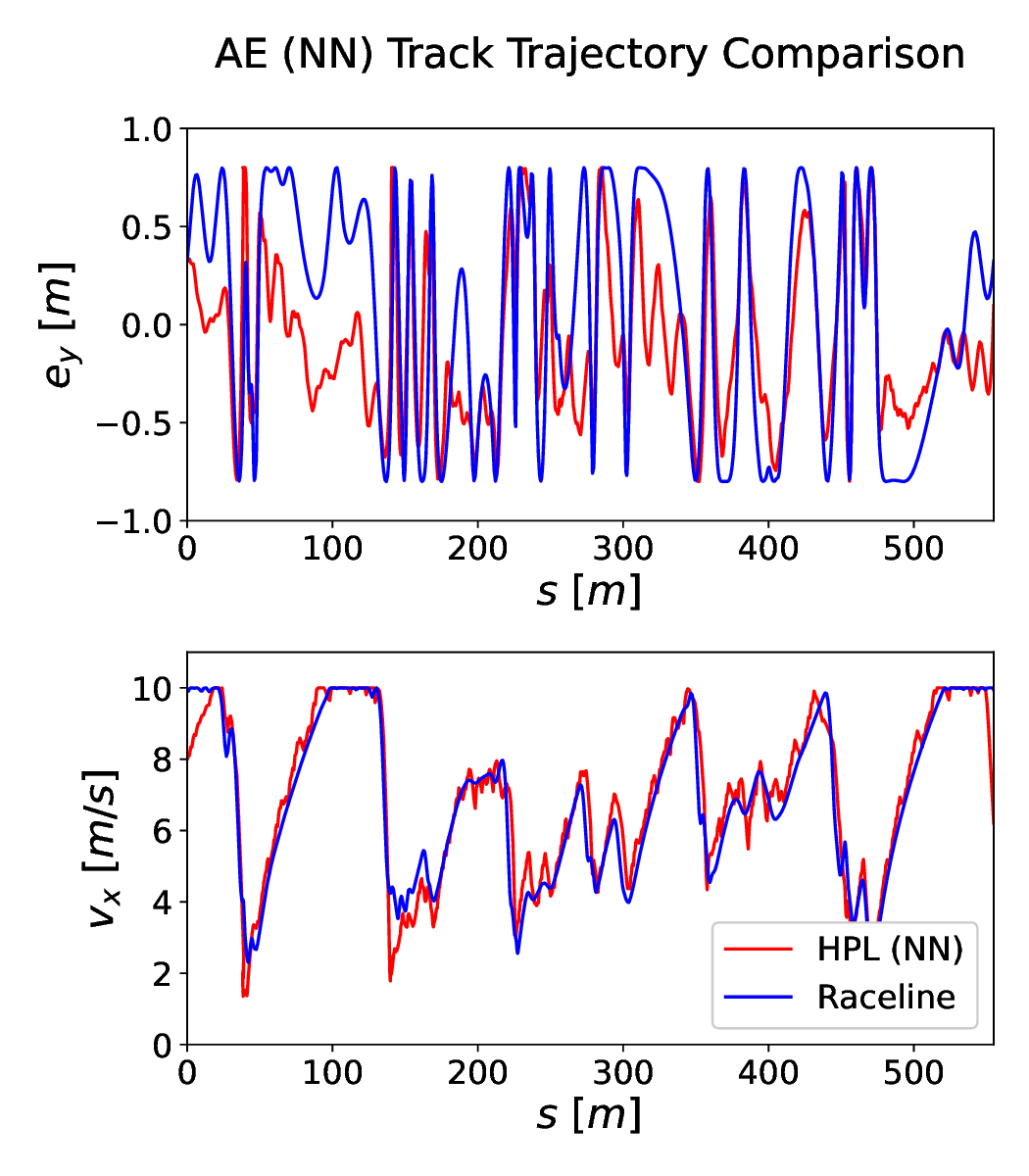}
    \caption{A strategy parameterized as a neural network can also capture the true raceline reasonably well, but with more errors in the predicted centerline deviation than the strategy GP.}
    \label{fig:nn_AE_traj}
\end{figure}

The resulting HPL trajectory on the AE track is shown and compared to the minimum-time trajectory in Figs.~\ref{fig:nn_AE_vel}-\ref{fig:nn_AE_traj}. 
Overall, the neural network strategy performs well, resulting in a lap time of 89.4 seconds, only slightly slower than the GP lap time (87.6 [s]) and the optimal lap time (85.1 [s]). 
This success validates that the HPL algorithm can easily be used with the designer's choice of strategy parameterizations, keeping in mind the considerations described in Sec.~\ref{sec:strategy_models}.

Interestingly, the $e_y$ and $v_x$ differences between the HPL trajectories (both of the GP and neural network) and the minimum time trajectories (as shown in Figs.~\ref{fig:nn_AE_traj} and \ref{fig:AE_trajcomp}) occur in similar regions of the track, including around the $s=100$ and $s=300$ meter marks, which are both long straightaway sections. 
The fact that both strategy parameterizations, the GP and the neural network, mischaracterized the race line in the same track section suggests that the training data could not be successfully exploited to race this section of the new track. 
Indeed, the ideal raceline along long straightaway sections is determined by the shape of the closest upcoming curve, which may lie beyond the environment forecast horizon and therefore unknown to the HPL control algorithm.
One approach to resolve this would be extending the environment forecast horizon, but this inevitably increases the computational complexity of both the offline training and online implementation of the HPL algorithm.

\section{Flappy Bird}\label{sec:flappybird}

\begin{figure}[t]
    \centering
    \includegraphics[width=0.8\columnwidth]{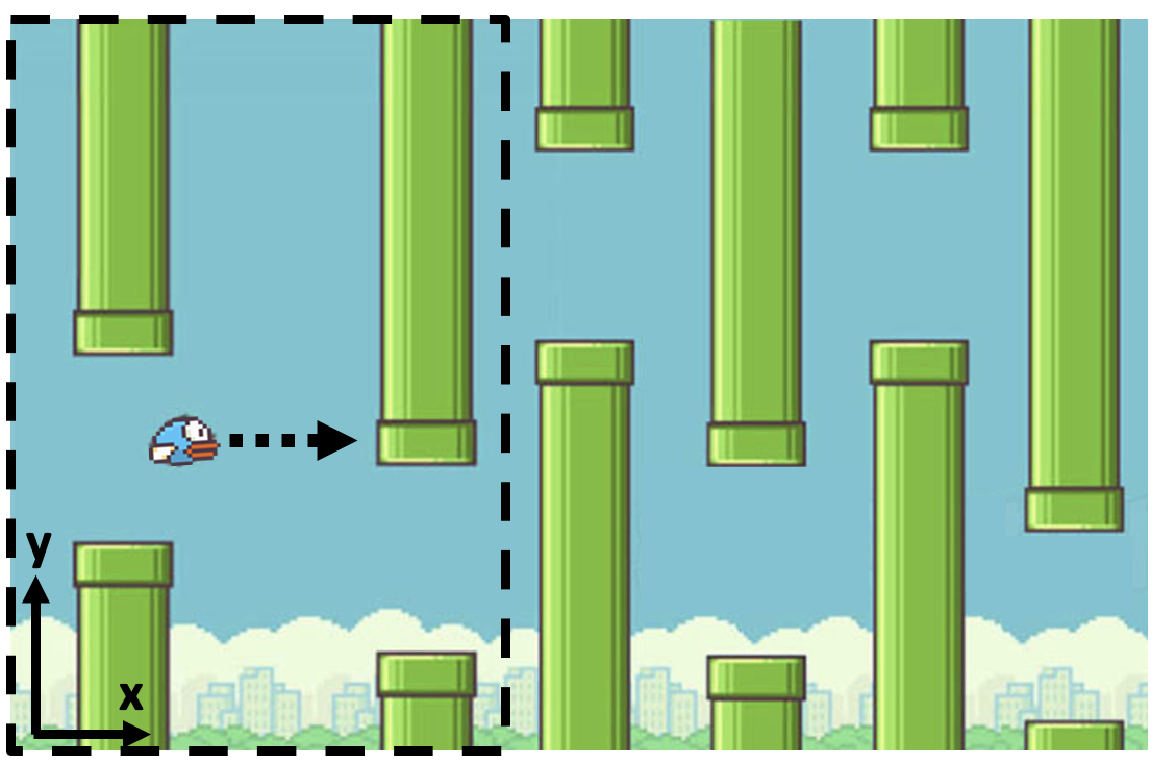}
    \caption{The goal of Flappy Bird is to guide the bird through a series of pipe obstacles. Only the pipes visible in the game screen (dashed rectangle) are visible to the bird at any time.}
    \label{fig:flappy_setup}
\end{figure}

We use Hierarchical Predictive Control to improve performance in the Flappy Bird computer game\footnote{\href{https://flappybird.io/}{\texttt{https://flappybird.io}}}. 
Here, the task is to steer a small bird around a series of pipe obstacles by controlling the timing of its wing flaps.
The pipe obstacles come in pairs from the bottom and top of the screen, leaving a gap for the bird to carefully fly through. As the bird moves through the task, it sees only a fixed distance ahead: the screen only the shows the two upcoming pairs of pipes.
The strategy behind Flappy Bird lies in planning short-term trajectories that are robust to randomness in the heights of the future pipe obstacles still hidden beyond the screen (see Fig.~\ref{fig:flappy_setup}). 

\subsection{System and Task Description}

We use a Flappy Bird simulator designed by Philip Zucker in \cite{zucker} for our experiments. Following Zucker's work, we model the bird as a simple three-state system with dynamics
\begin{align}\label{eq:flappydynamics}
    x_{k+1}&= x_k + 4, \\
    y_{k+1} &= y_k + {v_y}_k, \nonumber\\
    {v_y}_{k+1} &= {v_y}_k - 1 + 16u_k,\nonumber
\end{align}
where $x_k$ and $y_k$ are the horizontal and vertical position of the bird. The bird moves with constant horizontal velocity, while the vertical velocity ${v_y}_k$ is subject to a constant gravitational force and the wing flap control $u_k \in \{0,1 \}$, where $u_k=1$ implies a wing flap at time $k$. For each pair of pipe obstacles the bird flies through without crashing, a point is earned. %The goal in each Flappy Bird task is to earn as many points as possible before a crash ends the task. 

% The environment imposes state constraints in the form of pipes obstacles that must be avoided. 
% % The pipe obstacles come in pairs from the bottom and top of the screen, leaving a gap for the bird to fly through. While the vertical distance from the ground to the gap's center varies randomly between different pipe obstacle pairs, the vertical gap between a single pair of pipe obstacles is always $\Delta P$.
% % Because the widths of the pipes are fixed, the $i$th set of two pipe obstacles, $P^i$, can therefore be fully defined by the x- and y-coordinates of the top of the lower pipe, $x^{p,i}$ and $y^{p,i}$. 
% We denote the set of visible pipe obstacles (there are at most three) at time $k$ by $\mathcal{P}_k$,
% and the constraint that the bird may not crash into the pipes as
% \begin{align}
%     (x_k, y_k) \not\in \mathcal{P}_k.
% \end{align}

Different tasks $\mathcal{T}^i$ correspond to different sequences of pipe obstacles, described using ${\bf{\Theta}}^i$, where ${\bf{\Theta}}^i$ maps the bird's state $(x_k, y_k)$ to the set of pipe obstacles $\mathcal{P}_k$ visible at time $k$.

We choose the strategy states to be the vertical distances between the bird and the two closest upcoming pipe gaps:
\begin{align}
    \tilde{x}_k = [y^{p,1}_k - y_k, ~y^{p,2}_k - y_k],
\end{align}
where $y^{p,1}_k$ and $y^{p,2}_k$ are the heights of the two closest upcoming lower pipe obstacles. 
Note that because the bird has a constant horizontal velocity, we only need to consider strategy states that describe the bird's height - the bird's horizontal movement is predetermined, and not regulated by a strategy. 
%The strategy states measure how close the bird is to the upcoming obstacles at a given time step.
%Because the bird's dynamics (\ref{eq:flappydynamics}) have a constant horizontal velocity and the pipe obstacle height is completely random, an entirely robust safety controller is difficult to formulate. 
The safety control (\ref{eq:emergencycontrol}) is an MPC tracking the interpolated centerline of the visible pipe gaps. 

\subsection{Hierarchical Predictive Learning Results}
% Describe where the training data came from for the GP
We collect task executions that solve a series of 15 control tasks $\{\mathcal{T}^1, \dots, \mathcal{T}^{15}\}$, with each control task corresponding to a new instance of the Flappy Bird game. The executions are used to create training data for a GP, using an environment forecast of $N = 45$ and a control horizon of $T = 30$ time steps (with a sampling frequency of $30$ Hz.). This represents an environment forecast of two-thirds of the screen.
Because the horizontal velocity is fixed in these tasks, we only train a single strategy GP to predict the target vertical distance to the nearest pipe gap, $(y^{p,1}_{k+T} - y_{k+T})$. The GP training took $327$ seconds in GPyTorch with a 2017 MacBook Pro with 2.8 GHz Quad-Core Intel Core i7.

% Compare the numbers to the github controller, show a table
After training the GP model, the HPL algorithm is used to solve 50 different new Flappy Bird games. For comparison, we also evaluate a publicly available center-pipe tracking controller \cite{zucker} with the same control horizon on the same set of 50 games. 
Statistics of the scores earned by each controller during the 50 games are recorded below in Table \ref{sophisticatedtable}. 

\begin{center}
\begin{tabular}{ | l | c c c | } 
 \hline
  & Mean Score & Med. Score & Min. Score\\
 \hline
 Standard MPC  & 28 & 23 & 1\\ 
 HPL Control & 161 & 105 & 38\\ 
  pseudo-HPL & 74 & 46 & 10 \\ 
 \hline
\end{tabular}
\captionof{table}{The HPL controller is compared with a publicly available standard MPC controller  \cite{zucker}. In a trial of 50 tasks, the HPL controller significantly outperformed the standard controller.}\label{sophisticatedtable}
\end{center}

The HPL controller vastly outperforms the center-pipe tracking controller, with a mean score of 161 (versus 28). In fact, the HPL controller's minimum recorded score (38) is larger than the average score achieved by the standard controller. In the 50 evaluated games, there was only a single instance of the standard controller either outperforming or achieving the same score as the HPL controller on a specific task. 

HPL is able to outperform the standard controller by utilizing the safe set $\mathcal{X}_E$ to construct the terminal set at each time step, and ensure that the short-term trajectory plans lead to feasible solutions in future time steps. This is particularly important in the Flappy Bird task, because the pipe heights of two consecutive pipe obstacles are completely uncorrelated. Unlike in the autonomous racing task, where environments change gradually, the Flappy Bird controller needs to consider all possible future pipe heights at each time step.

\begin{figure}
    \centering
    \includegraphics[width=0.8\columnwidth]{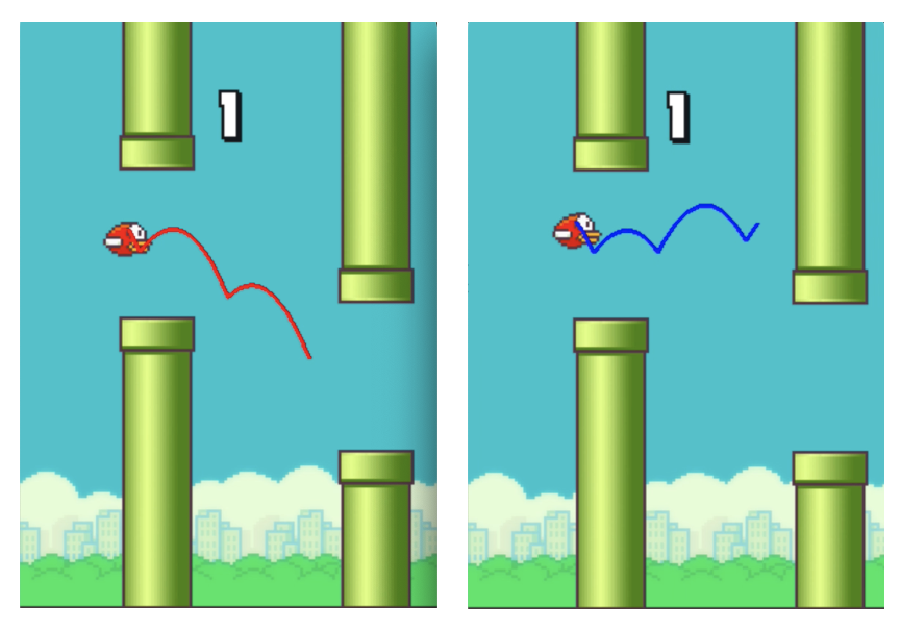}
    \caption{The open-loop trajectories of the HPL controller (left image, in red) and the standard MPC controller (right image, in blue) are compared. The HPL controller uses the pre-determined safety set $\mathcal{X}_E$ to plan trajectories that will be feasible regardless of the upcoming pipe obstacle height, resulting in a trajectory towards the center of the screen.}
    \label{fig:flappyControlCompare}
\end{figure}

Figure~\ref{fig:flappyControlCompare} compares the open-loop trajectory predictions for the HPL controller (left image, in red) and the standard controller (right image, in blue) while solving the same task. At the time step shown, the oncoming pipe is still beyond the control horizon (both controllers use the same control horizon). However, because the HPL controller plans trajectories that will be feasible for any kind of upcoming pipe obstacle, its open-loop trajectory aims downwards, towards the center of the screen. From this end position, the controller is more likely to find a feasible next trajectory regardless of the shape of the next obstacle.

In contrast, the standard controller continues to track the center of the pipe, without explicitly considering that the upcoming obstacle may be of a different shape. 
Because the control can only increase the rate of acceleration, whereas the bird's downward acceleration due to gravity is fixed, it is dangerous to plan for wing flaps close to still-unknown obstacles. Crashing into the side of pipe obstacle pipes as a result of too-close wing flaps was the most common failure mode of the standard controller in the games considered.
These crashes can only be prevented by using a standard controller if the control horizon $T$ is extended, but this will increase the computational complexity of the controller. 
Therefore it is clear that in Flappy Bird, the controller that plans ahead using strategies and a safety set outperforms one that does not.

% Show a side-by-side picture of two situations handled differently depending on which controller was used.

% Give a discussion on what parts were most valuable (how valuable was the GP really? Not very; it was mainly intersecting with a safe set in this case). Because it's just about *score*, which grows as much as it can as long as you remain feasible, there isn't really a difference here between safety and performance.
\subsection{Discussion}
It is worth noting that in this particular application, there is no difference between safety and performance. As long as the bird satisfies all environmental constraints, the score increases; there are no additional points for playing with a minimum number of wing flaps. Such an additional objective could, of course, be specified in an optimal control design, but it is not intrinsic to the game.

This direct correlation between Flappy Bird \textit{performance} and Flappy Bird \textit{safety} suggests that while the learned strategies may be useful for solving the task, they are not necessary in order to gain higher scores. In fact, a pseudo-HPL controller which only requires that the terminal state is in the safe set $\mathcal{X}_E$, without applying any learned strategies, also outperformed the standard controller (see Table \ref{sophisticatedtable}). 
This is likely due to the fact that a good approximation for the true $\mathcal{X}_E$ could be found and implemented in the safety controller alone. It is clear that in tasks with more complex dynamics or environments, where only rougher approximations of the safe set $\mathcal{X}_E$ can be determined, the additional incorporation of strategies to guide the system away from constraint violation will improve performance even further. Therefore, even when performance is directly linked to safety (as in Flappy Bird), all components of the HPL architecture play critical roles.

\section{Discussion}\label{sec:discussion}
% Next steps following that: Read through the examples (today)
% Read through everything, making comments (tomorrow)
% Make as many changes as possible by Friday (thursday, Friday)

The HPL algorithm provides a structured framework for solving tasks in new environments. Here, we would like to emphasize the benefits of the algorithm's key features.

\subsection{Incorporating Human Intuition}
Our proposed characteristics of strategy put forth in Sec.~\ref{sec:hpl} define strategies as mapping to target sets in a reduced dimension known as strategy space. As detailed in Sec.~\ref{sec:strategyfinder}, there are myriad ways of determining appropriate strategy states and inputs for a particular task, including using human intuition to select them. 

In some cases it can be tedious (or impossible) to hand-design features for control design; indeed this motivates much of the rise in ML for control. 
However, providing structured opportunities for incorporating human intuition wherever it may be available is still of great benefit. 
Only in rare cases will real-world systems be deployed with no understanding of their own dynamics or objectives in unknown and time-varying environments.
We believe it is much more realistic to have a control framework in mind that offers modular possibilities for human intuition to help shape control parameters whenever that intuition exists, such as the strategy states or forecast horizons in the HPL framework. 

\subsection{Ease of Interpretation}
One benefit of choosing intuitively meaningful strategy states is that it makes it easier to interpret the status of the learned strategies. 
If we can visualize the learned strategy sets and compare them to our human intuition about the task, we can easily understand if more training data is needed, or if the strategy mapping is logical. 
This analysis is an especially useful tool when trying to pinpoint a failure in the entire HPL control framework. 

\subsection{Using Data Safely}
The hierarchical approach of HPL allows for some separation between maximizing performance and ensuring feasibility: effective task performance is encouraged using learned strategies, and feasibility is ensured via the safety controller and safe set projections. 
(As discussed in Sec.~\ref{sec:flappybird}, there is some overlap between the two, especially in complex systems.)

In our proposed approach, these two hierarchical levels overlap with the algorithm's boundaries between data-driven and physics-based models. We use the physical model of the system in order to guarantee feasibility, and we use learning to represent complex objectives whose exact relationships are difficult characterize. 
This framework allows for responsible integration of data, using it to guide long-term behaviors rather than decide immediate control actions.

By varying the risk factor $\beta$ in our controller formulation, we were able to evaluate the benefits of using a physical model and safety controller to move from reduced-dimension strategy space to full-dimensional space. In all tested applications (see Sec.~\ref{sec:robot}-Sec.~\ref{sec:racing_taskdesc}), relying only on the data-driven strategy led to incomplete and infeasible task executions. These results emphasize the usefulness and importance of the hierarchical control framework, which allows for a structured combination of data-driven and physics-based methods.

\section{Conclusion}
A data-driven hierarchical predictive learning architecture for control in unknown environments is presented. 
The HPL algorithm uses stored executions that solve a variety of previous control tasks in order to learn a generalizable control strategy for new, unseen tasks. 
Based on a local description of the task environment, the learned control strategy proposes regions in the state space towards which to aim the system at each time step, and provides input constraints to guide the system evolution according to previous task solutions.
We prove that the resulting policy is guaranteed to be feasible for the new tasks, and evaluate the effectiveness of the proposed architecture in a simulation of a robotic path planning task. 
Our results confirm that HPL architecture is able to learn applicable strategies for efficient and safe execution of unseen tasks.

\appendices
\section*{Appendix}

\subsection{Additional Formula 1 Racing Plots}

The plots comparing the HPL controller trajectory to the optimal minimum-time trajectory for the remaining two test tracks (BE and US tracks) are shown here. As also shown in the figures in Sec.~\ref{sec:form1}, the HPL controller using short-sighted environmental forecasts in conjunction with strategy GPs results in a close-loop trajectory very similar to the raceline.

\begin{figure}
    \centering
    \includegraphics[width=0.9\columnwidth, trim = {7.5cm 0 1.2cm 0.2cm}, clip]{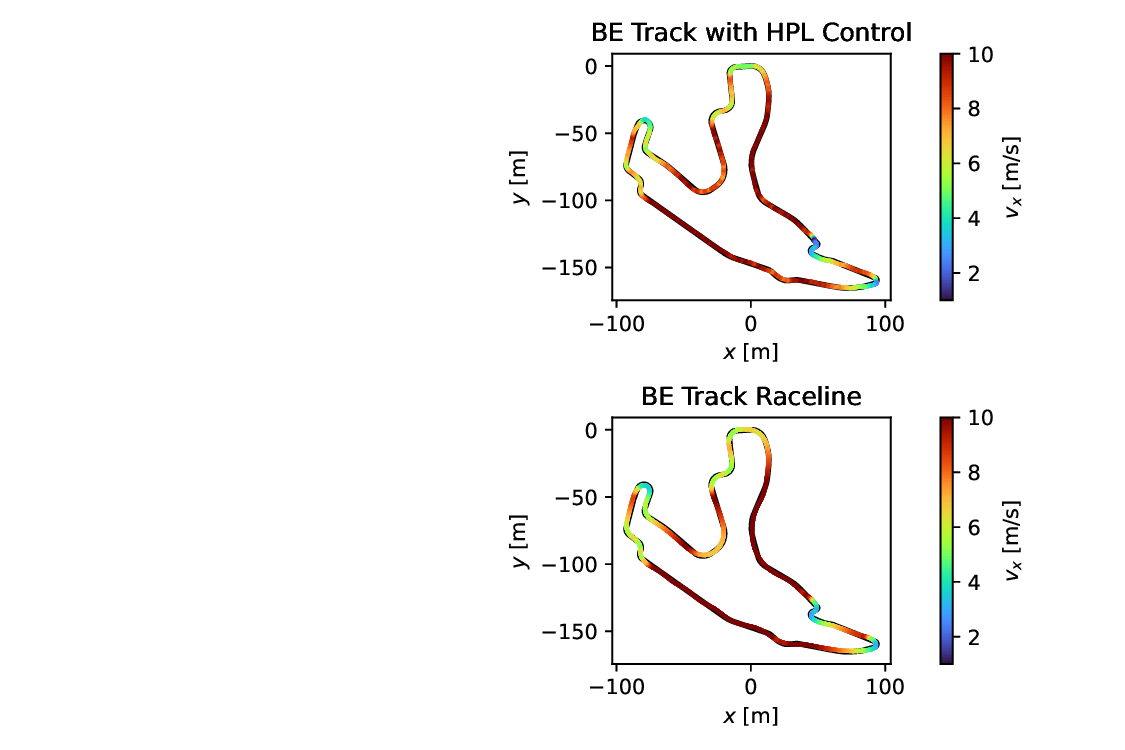}
    \caption{The HPL controller matches the speed profile and shape of the minimum time trajectory on the BE track.}
\end{figure}

\begin{figure}
    \centering
    \includegraphics[width=0.7\columnwidth, trim = {9.5cm 0 1.2cm 0.2cm}, clip]{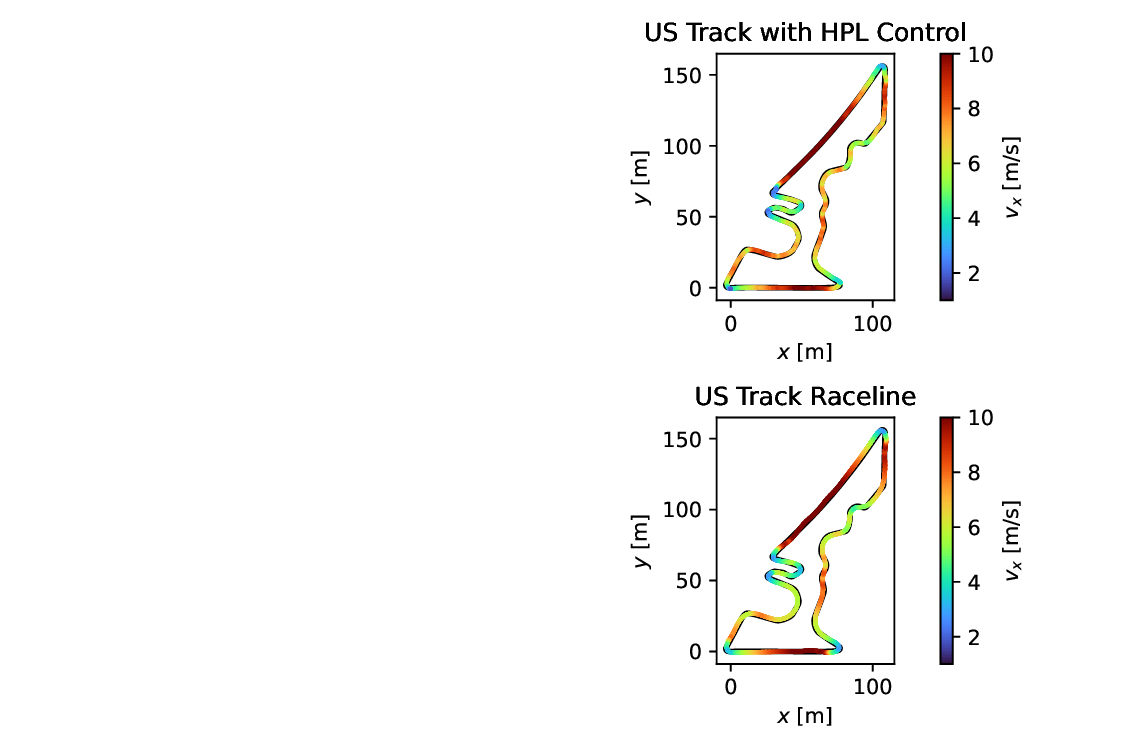}
    \caption{The HPL controller matches the speed profile and shape of the minimum time trajectory on the US track.}
\end{figure}

\bibliographystyle{IEEEtran}
\bibliography{IEEEabrv,main}

\begin{thebibliography}{10}
\providecommand{\url}[1]{#1}
\csname url@samestyle\endcsname
\providecommand{\newblock}{\relax}
\providecommand{\bibinfo}[2]{#2}
\providecommand{\BIBentrySTDinterwordspacing}{\spaceskip=0pt\relax}
\providecommand{\BIBentryALTinterwordstretchfactor}{4}
\providecommand{\BIBentryALTinterwordspacing}{\spaceskip=\fontdimen2\font plus
\BIBentryALTinterwordstretchfactor\fontdimen3\font minus
  \fontdimen4\font\relax}
\providecommand{\BIBforeignlanguage}[2]{{%
\expandafter\ifx\csname l@#1\endcsname\relax
\typeout{** WARNING: IEEEtran.bst: No hyphenation pattern has been}%
\typeout{** loaded for the language `#1'. Using the pattern for}%
\typeout{** the default language instead.}%
\else
\language=\csname l@#1\endcsname
\fi
#2}}
\providecommand{\BIBdecl}{\relax}
\BIBdecl

\bibitem{AA}
D.~A. Bristow, M.~Tharayil, and A.~G. Alleyne, ``A survey of iterative learning
  control,'' \emph{IEEE Control Systems Magazine}, vol.~26, no.~3, pp. 96--114,
  2006.

\bibitem{shen2019survey}
D.~Shen and X.~Li, ``A survey on iterative learning control with randomly
  varying trial lengths: Model, synthesis, and convergence analysis,''
  \emph{Annual Reviews in Control}, vol.~48, pp. 89--102, 2019.

\bibitem{ilcUAV}
\BIBentryALTinterwordspacing
J.~Dong and B.~He, ``Novel fuzzy pid-type iterative learning control for
  quadrotor uav,'' \emph{Sensors}, vol.~19, no.~1, 2019. [Online]. Available:
  \url{https://www.mdpi.com/1424-8220/19/1/24}
\BIBentrySTDinterwordspacing

\bibitem{armstrong2021multi}
A.~A. Armstrong and A.~G. Alleyne, ``A multi-input single-output iterative
  learning control for improved material placement in extrusion-based additive
  manufacturing,'' \emph{Control Engineering Practice}, vol. 111, p. 104783,
  2021.

\bibitem{hofer2019iterative}
M.~Hofer, L.~Spannagl, and R.~D’Andrea, ``Iterative learning control for fast
  and accurate position tracking with an articulated soft robotic arm,'' in
  \emph{2019 IEEE/RSJ International Conference on Intelligent Robots and
  Systems (IROS)}.\hskip 1em plus 0.5em minus 0.4em\relax IEEE, 2019, pp.
  6602--6607.

\bibitem{ketelhut2019iterative}
M.~Ketelhut, S.~Stemmler, J.~Gesenhues, M.~Hein, and D.~Abel, ``Iterative
  learning control of ventricular assist devices with variable cycle
  durations,'' \emph{Control Engineering Practice}, vol.~83, pp. 33--44, 2019.

\bibitem{rosolia2017autonomousrace}
U.~Rosolia, A.~Carvalho, and F.~Borrelli, ``Autonomous racing using learning
  model predictive control,'' in \emph{2017 American Control Conference
  (ACC)}.\hskip 1em plus 0.5em minus 0.4em\relax IEEE, 2017, pp. 5115--5120.

\bibitem{melanie}
J.~Kabzan, L.~Hewing, A.~Liniger, and M.~N. Zeilinger, ``Learning-based model
  predictive control for autonomous racing,'' \emph{IEEE Robotics and
  Automation Letters}, vol.~4, no.~4, pp. 3363--3370, 2019.

\bibitem{vermillion}
M.~K. {Cobb}, K.~{Barton}, H.~{Fathy}, and C.~{Vermillion}, ``Iterative
  learning-based path optimization for repetitive path planning, with
  application to 3-d crosswind flight of airborne wind energy systems,''
  \emph{IEEE Transactions on Control Systems Technology}, pp. 1--13, 2019.

\bibitem{vallon2019task}
C.~Vallon and F.~Borrelli, ``Task decomposition for iterative learning model
  predictive control,'' in \emph{2020 American Control Conference (ACC)}.\hskip
  1em plus 0.5em minus 0.4em\relax IEEE, 2020.

\bibitem{van2016optimality}
J.~van Zundert, J.~Bolder, and T.~Oomen, ``Optimality and flexibility in
  iterative learning control for varying tasks,'' \emph{Automatica}, vol.~67,
  pp. 295--302, 2016.

\bibitem{alleyne2010basis}
D.~J. Hoelzle, A.~G. Alleyne, and A.~J.~W. Johnson, ``Basis task approach to
  iterative learning control with applications to micro-robotic deposition,''
  \emph{IEEE Transactions on Control Systems Technology}, vol.~19, no.~5, pp.
  1138--1148, 2010.

\bibitem{rojas2017}
T.~Oomen and C.~R. Rojas, ``Sparse iterative learning control with application
  to a wafer stage: Achieving performance, resource efficiency, and task
  flexibility,'' \emph{Mechatronics}, vol.~47, pp. 134--147, 2017.

\bibitem{tomi2020neural}
D.~Zhang, Z.~Wang, and T.~Masayoshi, ``Neural-network-based iterative learning
  control for multiple tasks,'' \emph{IEEE Transactions on Neural Networks and
  Learning Systems}, 2020.

\bibitem{patan2020neural}
K.~Patan and M.~Patan, ``Neural-network-based iterative learning control of
  nonlinear systems,'' \emph{ISA transactions}, vol.~98, pp. 445--453, 2020.

\bibitem{transferlearning}
L.~Torrey and J.~Shavlik, ``Transfer learning,'' \emph{Handbook of Research on
  Machine Learning Applications}, 01 2009.

\bibitem{weiss2016survey}
K.~Weiss, T.~M. Khoshgoftaar, and D.~Wang, ``A survey of transfer learning,''
  \emph{Journal of Big data}, vol.~3, no.~1, p.~9, 2016.

\bibitem{zhuang2019comprehensive}
F.~Zhuang, Z.~Qi, K.~Duan, D.~Xi, Y.~Zhu, H.~Zhu, H.~Xiong, and Q.~He, ``A
  comprehensive survey on transfer learning,'' \emph{arXiv preprint
  arXiv:1911.02685}, 2019.

\bibitem{finnMetaRL2018}
I.~Clavera, A.~Nagabandi, R.~S. Fearing, P.~Abbeel, S.~Levine, and C.~Finn,
  ``Learning to adapt in dynamic, real-world environments through
  meta-reinforcement learning,'' \emph{arXiv preprint arXiv:1803.11347}, 2018.

\bibitem{levine2018hierarchical}
O.~Nachum, S.~S. Gu, H.~Lee, and S.~Levine, ``Data-efficient hierarchical
  reinforcement learning,'' in \emph{Advances in neural information processing
  systems}, 2018, pp. 3303--3313.

\bibitem{bertsimas2018voice}
D.~Bertsimas and B.~Stellato, ``The voice of optimization,'' 2018.

\bibitem{hewing2019cautious}
L.~Hewing, J.~Kabzan, and M.~N. Zeilinger, ``Cautious model predictive control
  using gaussian process regression,'' \emph{IEEE Transactions on Control
  Systems Technology}, 2019.

\bibitem{stolle}
M.~Stolle and C.~Atkeson, ``Finding and transferring policies using stored
  behaviors,'' \emph{Autonomous Robots}, vol.~29, no.~2, pp. 169--200, 2010.

\bibitem{berenson}
D.~Berenson, P.~Abbeel, and K.~Goldberg, ``A robot path planning framework that
  learns from experience,'' in \emph{2012 IEEE International Conference on
  Robotics and Automation}.\hskip 1em plus 0.5em minus 0.4em\relax IEEE, 2012,
  pp. 3671--3678.

\bibitem{pereida2018data}
K.~Pereida, M.~K. Helwa, and A.~P. Schoellig, ``Data-efficient multirobot,
  multitask transfer learning for trajectory tracking,'' \emph{IEEE Robotics
  and Automation Letters}, vol.~3, no.~2, pp. 1260--1267, 2018.

\bibitem{zhi2019octnet}
W.~Zhi, T.~Lai, L.~Ott, G.~Francis, and F.~Ramos, ``Octnet: Trajectory
  generation in new environments from past experiences,'' 2019.

\bibitem{probabilisticprimitives}
A.~Paraschos, G.~Neumann, and J.~Peters, ``A probabilistic approach to robot
  trajectory generation,'' in \emph{13th IEEE-RAS International Conference on
  Humanoid Robots (Humanoids)}.\hskip 1em plus 0.5em minus 0.4em\relax IEEE,
  2013, pp. 477--483.

\bibitem{fitzgerald}
T.~Fitzgerald, E.~Short, A.~Goel, and A.~Thomaz, ``Human-guided trajectory
  adaptation for tool transfer,'' in \emph{Proceedings of the 18th
  International Conference on Autonomous Agents and MultiAgent Systems}, ser.
  AAMAS '19, 2019, pp. 1350--1358.

\bibitem{6}
M.~Stolle and C.~Atkeson, ``Finding and transferring policies using stored
  behaviors,'' \emph{Autonomous Robots}, vol.~29, no.~2, pp. 169--200, 2010.

\bibitem{liu2009standing}
C.~Liu and C.~G. Atkeson, ``Standing balance control using a trajectory
  library,'' in \emph{2009 IEEE/RSJ International Conference on Intelligent
  Robots and Systems}.\hskip 1em plus 0.5em minus 0.4em\relax Citeseer, 2009,
  pp. 3031--3036.

\bibitem{7}
Y.~Tassa, T.~Erez, and W.~D. Smart, ``Receding horizon differential dynamic
  programming,'' in \emph{Advances in neural information processing systems},
  2008, pp. 1465--1472.

\bibitem{vallon2020data}
C.~Vallon and F.~Borrelli, ``Data-driven hierarchical predictive learning in
  unknown environments,'' \emph{arXiv preprint arXiv:2005.05948}, 2020.

\bibitem{bertsekasagg}
D.~P. Bertsekas, ``Feature-based aggregation and deep reinforcement learning: A
  survey and some new implementations,'' \emph{arXiv preprint
  arXiv:1804.04577}, 2018.

\bibitem{learningGaussProcess}
A.~Jain, T.~X. Nghiem, M.~Morari, and R.~Mangharam, ``Learning and control
  using gaussian processes: Towards bridging machine learning and controls for
  physical systems,'' in \emph{Proceedings of the 9th ACM/IEEE International
  Conference on Cyber-Physical Systems}, ser. ICCPS ’18, 2018, p. 140–149.

\bibitem{klenske2015gaussian}
E.~D. Klenske, M.~N. Zeilinger, B.~Sch{\"o}lkopf, and P.~Hennig, ``Gaussian
  process-based predictive control for periodic error correction,'' \emph{IEEE
  Transactions on Control Systems Technology}, vol.~24, no.~1, pp. 110--121,
  2015.

\bibitem{kocijan2016modelling}
J.~Kocijan, \emph{Modelling and control of dynamic systems using Gaussian
  process models}.\hskip 1em plus 0.5em minus 0.4em\relax Springer.

\bibitem{bujarbaruahAdapFIR}
M.~Bujarbaruah, X.~Zhang, and F.~Borrelli, ``Adaptive {MPC} with chance
  constraints for {FIR} systems,'' in \emph{2018 Annual American Control
  Conference (ACC)}, June 2018, pp. 2312--2317.

\bibitem{nandeshwar2006models}
A.~R. Nandeshwar, ``Models for calculating confidence intervals for neural
  networks,'' 2006.

\bibitem{ott2018analyzing}
M.~Ott, M.~Auli, D.~Grangier, and M.~Ranzato, ``Analyzing uncertainty in neural
  machine translation,'' 2018.

\bibitem{extraLayers1}
A.~{Khosravi}, S.~{Nahavandi}, D.~{Creighton}, and A.~F. {Atiya}, ``Lower upper
  bound estimation method for construction of neural network-based prediction
  intervals,'' \emph{IEEE Transactions on Neural Networks}, vol.~22, no.~3, pp.
  337--346, 2011.

\bibitem{papadopoulos2001confidence}
G.~Papadopoulos, P.~J. Edwards, and A.~F. Murray, ``Confidence estimation
  methods for neural networks: A practical comparison,'' \emph{IEEE
  transactions on neural networks}, vol.~12, no.~6, pp. 1278--1287, 2001.

\bibitem{towardsBetter}
V.~T. {Vasudevan}, A.~{Sethy}, and A.~R. {Ghias}, ``Towards better confidence
  estimation for neural models,'' in \emph{ICASSP 2019 - 2019 IEEE
  International Conference on Acoustics, Speech and Signal Processing
  (ICASSP)}, 2019, pp. 7335--7339.

\bibitem{cortes2018deep}
I.~Cort{\'e}s-Ciriano and A.~Bender, ``Deep confidence: a computationally
  efficient framework for calculating reliable prediction errors for deep
  neural networks,'' \emph{Journal of chemical information and modeling},
  vol.~59, no.~3, pp. 1269--1281, 2018.

\bibitem{robey2020model}
A.~Robey, H.~Hassani, and G.~J. Pappas, ``Model-based robust deep learning,''
  \emph{arXiv preprint arXiv:2005.10247}, 2020.

\bibitem{data_invariant_sets}
A.~{Chakrabarty}, A.~{Raghunathan}, S.~{Di Cairano}, and C.~{Danielson},
  ``Data-driven estimation of backward reachable and invariant sets for
  unmodeled systems via active learning,'' in \emph{2018 IEEE Conference on
  Decision and Control (CDC)}, 2018, pp. 372--377.

\bibitem{wang2020datadriven}
Z.~Wang and R.~M. Jungers, ``Data-driven computation of invariant sets of
  discrete time-invariant black-box systems,'' 2020.

\bibitem{ozay2018}
Y.~{Chen}, H.~{Peng}, J.~{Grizzle}, and N.~{Ozay}, ``Data-driven computation of
  minimal robust control invariant set,'' in \emph{2018 IEEE Conference on
  Decision and Control (CDC)}, 2018, pp. 4052--4058.

\bibitem{ugo2020unified}
U.~Rosolia, A.~Singletary, and A.~D. Ames, ``Unified multi-rate control: from
  low level actuation to high level planning,'' 2020.

\bibitem{robothard3}
M.~Spong and M.~Vidyasagar, \emph{Robot Dynamics And Control}, 01 1989.

\bibitem{rosolia2016learning}
U.~Rosolia and F.~Borrelli, ``Learning model predictive control for iterative
  tasks. a data-driven control framework,'' \emph{IEEE Transactions on
  Automatic Control}, vol.~63, no.~7, pp. 1883--1896, 2017.

\bibitem{ugoproof}
\BIBentryALTinterwordspacing
------, ``Learning model predictive control for iterative tasks,'' \emph{CoRR},
  vol. abs/1609.01387, 2016. [Online]. Available:
  \url{http://arxiv.org/abs/1609.01387}
\BIBentrySTDinterwordspacing

\bibitem{rajamani2011vehicle}
R.~Rajamani, \emph{Vehicle dynamics and control}.\hskip 1em plus 0.5em minus
  0.4em\relax Springer Science \& Business Media, 2011.

\bibitem{belien}
\BIBentryALTinterwordspacing
J.~Beliën, ``jbelien/f1-circuits.'' [Online]. Available:
  \url{https://github.com/jbelien/F1-Circuits}
\BIBentrySTDinterwordspacing

\bibitem{zucker}
\BIBentryALTinterwordspacing
P.~Zucker, ``philzook58/flappybird-mpc.'' [Online]. Available:
  \url{https://github.com/philzook58/FlapPyBird-MPC}
\BIBentrySTDinterwordspacing

\end{thebibliography}

\begin{IEEEbiography}[{\includegraphics[width=1in,height=1.25in,clip,keepaspectratio]{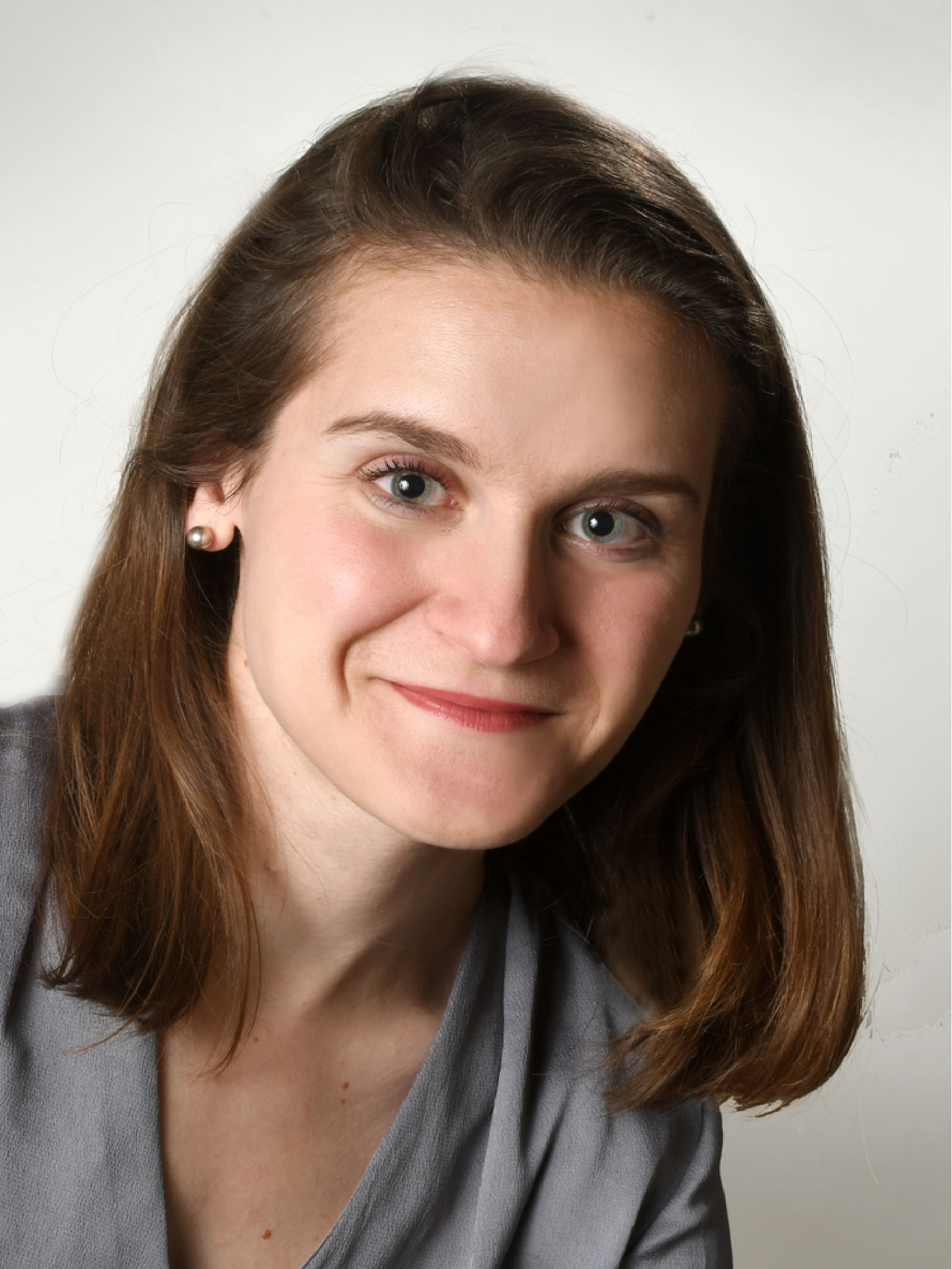}}]{Charlott S. Vallon} received the B.S. degree in mechanical engineering from the University of California at Berkeley, Berkeley, CA, USA, in 2015 and the M.S. degree from ETH-Zürich, Zürich, Switzerland, in 2017. She is currently a Ph.D. candidate at the University of California at Berkeley and a GMSE Fellow.

She was a Visiting Scholar with the University of California at Berkeley from 2016 to 2017. 
She was a Machine Learning Intern with Autodesk in San Francisco, CA, USA in 2016 and a Visiting GMSE Fellow Researcher at the National Institute of Standards and Technology in Gaithersburg, MD, USA in 2019. Her current research interests include hierarchical control, safe control in uncertain environments, and combining data-driven methods with traditional control theory.
\end{IEEEbiography}

\begin{IEEEbiography}[{\includegraphics[height=1.25in,clip,keepaspectratio]{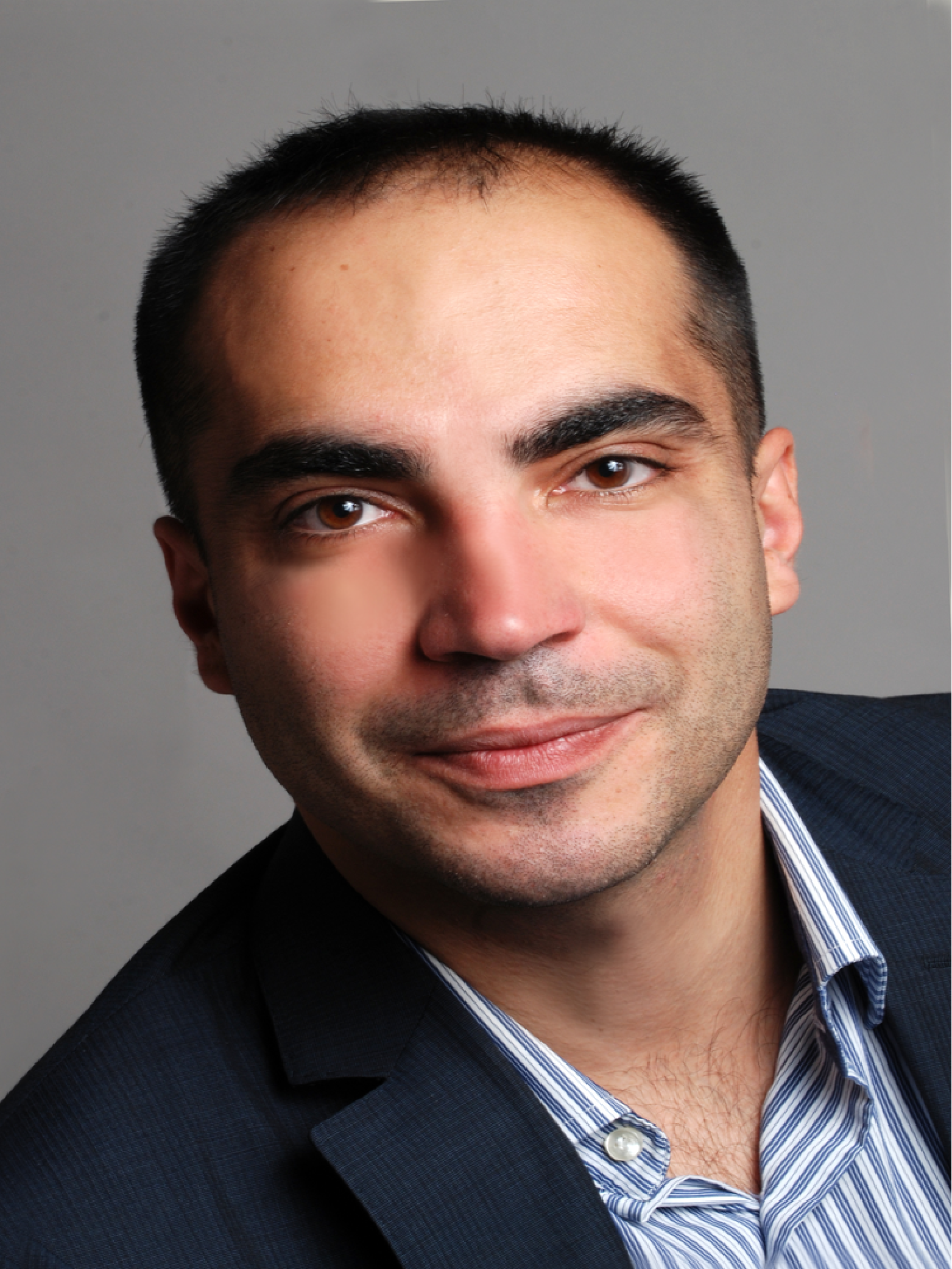}}]{Francesco Borrelli} received the Laurea degree in
computer science engineering from the University
of Naples Federico II, Naples, Italy, in 1998,
and the Ph.D. degree from ETH-Zürich, Zürich,
Switzerland, in 2002.

He is currently a Professor with
the Department of Mechanical Engineering, University of California, Berkeley, CA, USA. He
is the author of more than 100 publications in
the field of predictive control and author of the
book Constrained Optimal Control of Linear and
Hybrid Systems (Springer Verlag). His research interests include constrained
optimal control, model predictive control and its application to advanced
automotive control and energy efficient building operation.

Dr. Borrelli was the recipient of the 2009 National Science Foundation
CAREER Award and the 2012 IEEE Control System Technology Award.
In 2008, he was appointed the chair of the IEEE technical committee on
automotive control.
\end{IEEEbiography}

\end{document}